\documentclass[12pt]{article}
\usepackage[final]{graphicx}
\usepackage{amsmath}
\usepackage{amssymb}
\usepackage{xcolor}
\usepackage{tensor}
\usepackage{braket}
\usepackage{units}

\usepackage{amssymb}
\newcounter{comment}

{\refstepcounter{comment}%
\begin{quote}
\ttfamily\small$\blacksquare$ \textbf{\underline{Comment} $\sharp$\thecomment:}}%
{\end{quote}}

{
\begin{quote}
\ttfamily\small$\blacktriangleright$ \textbf{\underline{Reply} $\sharp$\thecomment:}}%
{\end{quote}}

\newcommand{\xB}{x_{\rm B}}

\newcommand{\GeV}{{\rm GeV}}

\def\muFgpd{\relax\ifmmode\mu_\text{F,GPD}^2\else{$\mu_\text{F,GPD}^2${ }}\fi}
\def\muFda{\relax\ifmmode\mu_\text{F,DA}^2\else{$\mu_\text{F,DA}^2${ }}\fi}
\def\muO{\relax\ifmmode{\mu_{0}^{2}}\else{$\mu_{0}^{2}${ }}\fi}
\def\Mev{\relax\ifmmode{\text{MeV}}\else{MeV{ }}\fi}

\def\Li{\relax\ifmmode{\text{Li}_{2}}\else{Li$_2${ }}\fi}

\font\cmss=cmss12 
\def\1{\hbox{{1}\kern-.25em\hbox{l}}}
\def\bfZ{\relax{\hbox{\cmss Z\kern-.4em Z}}}
\newcommand{\be}{\begin{eqnarray}}
\newcommand{\ee}{\end{eqnarray}}
\newcommand{\ba}{\begin{array}}
\newcommand{\ea}{\end{array}}

\newcommand{\bi}{\begin{itemize}}
\newcommand{\ei}{\end{itemize}}

\begin{document}

\begin{titlepage}

\centerline{\large \bf Dual parametrization of generalized parton distributions}
\centerline{\large \bf in two equivalent representations}

\vspace{5mm}

\centerline{D.~M\"{u}ller$^a$,
            M.~V.~Polyakov$^{b, \,c}$ and
            K.~M.~Semenov-Tian-Shansky$^d$}

\vspace{4mm}

\centerline{\it $^a$Department of Physics, University of Cape Town}
\centerline{\it  ZA-7701  Rondebosch, South Africa}

\vspace{1mm}

\centerline{\it $^b$Institut f\"ur Theoretische
Physik II, Ruhr-Universit\"at Bochum}
\centerline{\it  DE-44780 Bochum, Germany}

\vspace{1mm}

\centerline{\it $^c$ Petersburg Nuclear Physics Institute, Gatchina}
\centerline{\it RU-188300, St. Petersburg, Russia}

\vspace{1mm}

\centerline{\it $^d$ IFPA, d{\'e}partement AGO, Universit{\' e} de Li{\`e}ge}
\centerline{\it BE-4000 Li{\` e}ge, Belgium}

\vspace{5mm}

\centerline{\bf Abstract}\vspace{2mm}

\noindent
The dual parametrization and the Mellin-Barnes integral approach represent two frameworks
for handling the double partial wave expansion of generalized parton distributions (GPDs) in
the conformal partial waves and in the $t$-channel
${\rm SO}(3)$
partial waves.
Within the dual parametrization framework, GPDs are represented
as integral convolutions of forward-like functions  whose
Mellin moments generate the conformal moments of GPDs.
The Mellin-Barnes integral approach is based on the analytic
continuation of the GPD conformal moments to the complex values
of the conformal spin. GPDs are then represented as the Mellin-Barnes-type integrals in the complex conformal spin plane.
In this paper we explicitly show the equivalence
of these two independently developed GPD representations. Furthermore, we clarify the notions of the
$J=0$
fixed pole and the
$D$-form factor.
We also provide some insight into GPD modeling and map the phenomenologically successful
Kumeri{\v c}ki-M{\"u}ller
GPD model to the dual parametrization framework by presenting the set of the corresponding forward-like functions.
We also build up the reparametrization procedure allowing to recast the double distribution  representation
of GPDs in the Mellin-Barnes integral framework and present the explicit formula for  mapping
double distributions into the space of double partial wave amplitudes with complex conformal spin.

\noindent

\vspace{0.5cm}

\noindent

\vspace*{12mm}


\end{titlepage}

\tableofcontents
\newpage

\section{Introduction}
The concept of generalized parton distributions (GPDs) introduced two decades ago
\cite{Mueller:1998fv,Ji:1996ek,Radyushkin:1996nd,Ji:1996nm,Collins:1996fb}
proved to be successful for addressing the issue of hadronic structure in terms of the fundamental degrees
of freedom of QCD. Considerable amount of experimental information acquired  during the last years inspired
further efforts for better understanding of both theoretical and experimental aspects of the GPD framework
(see reviews in Refs.
\cite{Goeke:2001tz,Diehl:2003ny,Belitsky:2005qn,
Mueller:2014hsa,Muller:2014tqa}).
Nowadays, dedicated experiments aiming on the detailed studies of hard exclusive reactions (such as
the Deeply Virtual Compton Scattering (DVCS) and the Deeply Virtual Meson Production (DVMP)) admitting
description within the  GPD formalism constitute a significant part of the research programs of several
existing (JLab, COMPASS) and future (EIC, \={P}ANDA @ GSI-FAIR, J-PARC) experimental facilities. This makes hope
that  precise experimental data will be available for scrutinizing GPDs in the future.
However, the direct extraction of GPDs from the observable quantities represents a formidable task.
Indeed, GPDs are intricate functions of the longitudinal momentum fraction of partons
($x$),
skewness parameter
($\eta$),
the momentum transfer squared
($t$),
and the factorization scale
($\mu$).
Moreover, GPDs enter observable quantities as integral convolutions over
$x$
with hard scattering kernels given by the appropriate partonic propagators. The direct deconvolution
for recovering GPDs from observables turns to be impracticable. Indeed, the leading order (LO) DVCS and DVMP
observables probe GPDs only on the so-called cross-over line
$x=\eta$.
Outside the cross-over line GPDs can be constrained only implicitly via the evolution effects.
A small lever arm in the experimentally available virtuality of the hard probes
$Q^2$
causes further embarrassment for this straightforward approach.

Therefore, a more realistic strategy for extracting GPDs from the data relies on employing of phenomenologically
motivated GPD representations and consistent GPD fitting
procedures for the whole set of observable quantities.
The clue for building up a valid phenomenological representation for GPDs is provided by implementation of the
non-trivial requirements (forward limit, polynomiality and support properties, hermiticity, positivity
constraints, evolution properties, {\it etc.}) following from the fundamental properties of the underlying
quantum field theory.

It is worth emphasizing that, as long as the basic theoretical requirements are satisfied, all GPD representations
present the same field theoretical object. Therefore, in principle, it should be possible to map a GPD within one
representation to that in a different representation, although explicitly working out such relations may
sometimes be mathematically cumbersome. This generally makes the popular question ``Which GPD representation is better?''
meaningless. Instead, one may hope to get an additional insight of GPDs and their physical
interpretation by comparing the manifestation of GPD properties within different representations.

Historically, one of the first parametrizations for GPDs suitable for phenomenological applications was based
on the double distribution (DD) representation of GPDs, introduced in \cite{Mueller:1998fv,Radyushkin:1997ki}.
The DD representation of GPDs arises
directly from  diagrammatical considerations and inherits most of the basic field theoretic
requirements for GPDs. It provides an elegant way to implement both polynomiality and
support properties as well as the forward limit constraint. Particular GPD models designed within this framework are
based on the Radyushkin Double Distribution Ansatz (RDDA)
\cite{Radyushkin:1998bz,Radyushkin:1998es,Musatov:1999xp}.
The popular Vanderhaeghen-Guichon-Guidal (VGG) code
\cite{Vanderhaeghen:1999xj}
for the DVCS and DVMP observables
implements the RDDA
model specified in \cite{Goeke:2001tz,Guidal:2004nd}.
The VGG code predictions and also those of the related Goloskokov-Kroll model
\cite{Goloskokov:2005sd,Goloskokov:2007nt}
have shown some success in the description of the
available data.
In our days, the RDDA still remains extremely popular and is also employed within more advanced GPD models
\cite{Radyushkin:2011dh,Radyushkin:2013hca},
which exploit the ambiguity
\cite{Belitsky:2000vk,Teryaev:2001qm}
of the DD representation.

Nevertheless, it is important to emphasize that the RDDA was originally designed as a toy model
\cite{Musatov:1999xp},
and thus it comes as no surprise that it possesses some serious drawbacks.
\bi
\item In particular, the functional form of the RDDA is not stable under the QCD evolution making the proper implementation
of the evolution effects arduous.
\item Also, already for one decade, it was recognized that the RDDA lacks flexibility, and, as a consequence,
most probably fails to describe the available H1 and ZEUS data for the  DVCS cross section at small values of
$\xB$ in the leading order of perturbation theory \cite{Freund:2002qf}.
\ei
Therefore, on no account the RDDA  should be taken as the only opportunity for phenomenological applications.

 The alternative strategy for building up of a GPD representation resides on the expansion of GPDs over a suitable
orthogonal polynomial basis in order to achieve the separation of certain variable dependence. The appealing
possibility is to perform the expansion of GPDs over the conformal partial wave  basis, which ensures the
diagonalization of the leading order (LO) evolution operator%
\footnote{Also the possible fundamental importance of the GPD parameterization
in terms of conformal partial waves, was recently emphasized in the context
of gravity dual description for the DVCS (see
\cite{Nishio:2012qh,Nishio:2014eua}
and references therein). }.
This inspired several authors to set up various versions of
the conformal partial wave expansion for GPDs. Nowadays two main versions of such GPD representations are utilized:
\begin{itemize}
\item The approach
\cite{Mueller:2005ed,Kirch:2005tt,Manashov:2005xp}
based on the Mellin-Barnes integral techniques. It deals with the analytic continuation of conformal
moments and of conformal partial waves to the complex values of conformal spin.
The closed expression for GPDs is then worked out
by trading the conformal partial wave expansion for the Mellin-Barnes type integral by means of the Sommerfeld-Watson
transformation.

\item The approach
\cite{Shuvaev:1999ce,Shuvaev:1999fm,Noritzsch:2000pr}
based on the Shuvaev-Noritzsch transform. In this approach one is dealing with the so-called
forward-like functions of an auxiliary momentum-fraction-type variable. The Mellin moments of these forward-like functions generate the conformal
moments of GPDs. GPDs are then given by convolutions of the forward-like functions with certain integral
kernels.
\end{itemize}

To proceed with the conformal partial wave expansion of GPDs it turns out
extremely instructive to further expand the conformal moments over the basis of
the $t$-channel
${\rm SO}(3)$
rotation group partial waves.
In the context of the Shuvaev-Noritzsch transform techniques the resulting GPD representation
is known as the dual parametrization of GPDs
\cite{Polyakov:2002wz,Polyakov:2008aa}.
Within the Mellin-Barnes integral approach, this kind of double partial wave expansion
is referred in the literature as the
${\rm SO}(3)$
partial wave expansion
\cite{Kumericki:2006xx}.
Each version of the formalism employs a rather intricate mathematical apparatus.
Therefore, some care is needed for the correct mathematical treatment, and surely some early
results/claims required refinement in our days.

The Mellin-Barnes integral approach
was successfully employed in a global GPD fitting framework.
Using a simple phenomenological Ansatz for the corresponding partial wave amplitudes
allows to construct a flexible Kumeri{\v c}ki-M{\"u}ller (KM)
GPD model
\cite{Kumericki:2006xx,Kumericki:2007sa,Mueller:2013caa}
that provides a good description of the world sets of unpolarized DVCS data \cite{Kumericki:2009uq}, including small-$\xB$ DVMP data \cite{Lautenschlager:2013uya}.
This  model provides also a reasonable description of the polarized DVCS data \cite{Kumericki:2013br,Seder:2014oaa}.

The early attempts
\cite{Guzey:2005ec,Guzey:2006xi,Polyakov:2008xm}
to apply  the dual parametrization approach for data analysis
were based on the so-called `minimalist' model which was later found to be based on too
strong theoretical assumptions. The corresponding data analysis was
much less consecutive and only partially consistent
\cite{Kumericki:2007mh,Guzey:2008ys}.
This is, however, not related to any fundamental drawback of the dual parametrization formalism.
In fact, both the dual parametrization and  the
${\rm SO}(3)$
partial wave expansion within the Mellin-Barnes integral approach  allow to set up very flexible GPD models in a trivial way
and are expected to be capable to describe the DVCS and DVMP data in a  global analysis at  next-to-leading order.
Therefore, we think that it is instructive to show in details that these two versions
of the conformal partial wave expansion formalism are fully equivalent.
This is the main goal of the present paper.

The paper is organized as follows.
In Sec.~\ref{sec:relations} we introduce our notations and present the details of the conformal 
partial wave expansion  for quark GPDs.
In Sec.~\ref{sec:DualParametrization} we briefly review the dual parametrization approach and
present a new derivation of the explicit form of the dual parametrization convolution kernels based on the
Schl{\" a}fli contour integral representation for the conformal partial waves.
In Sec.~\ref{sec:MellinBarnesSO(3)}, on general mathematical ground,
we show  the complete equivalence of the dual parametrization
and of the Mellin-Barnes ${\rm SO}(3)$ partial wave expansion approaches.
This equivalence is further illustrated by treating several important special cases in
Sec.~\ref{sec:SpecialCases}.
In Sec.~\ref{sec:Amplitude} we consider the elementary LO amplitude and discuss the manifestation of the
$J=0$
fixed pole in DVCS.
In Sec.~\ref{sec:example}
we provide concrete model examples
and recast the  phenomenological KM  model for GPDs
into the dual parametrization framework.
In Sec.~\ref{sec:dPWAsfromDDs} we evaluate conformal GPD moments, expanded in terms of
${\rm SO}(3)$ partial waves,
from double distributions. Finally, we give a summary and report on the status of the formalism.

\section{Conformal partial wave expansions of GPDs}
\label{sec:relations}

\subsection{Notations and conventions}

The conformal partial wave expansion (PWE) of GPDs deals with  partial
waves (PWs) that are labeled by the complex conformal spin, which characterizes the irreducible multiplets
of appropriate conformal operators.
We refer the reader {\it e.g.} to Ref.~\cite{Braun:2003rp}
for a review of group theoretical aspects of the conformal PWE.
In the present section we specify our set of conventions and notations for
the conformal basis and summarize the details of the conformal PWE for quark GPDs.

For a particular quark flavor $q$ a generic quark GPD
$F(x,\eta,t)= \{ H, \, E,\, \tilde{H}, \, \tilde{E}\} (x,\eta,t)$
with the support
$x\in [-1,1]$
can be decomposed  into
a quark GPD $F^q$
\be
F^q(x,\eta,t)= F(x\ge -|\eta|,\eta,t)
\label{q-part_GPD}
\ee
with the support
$x\in [-|\eta|,1]$
and an antiquark GPD $F^{\bar{q}}$
\be
F^{\bar{q}}(x,\eta,t)=-F(-x\le |\eta|,\eta,t)
\ee
with the support
$x\in[-1,|\eta|]$.
Both quark and antiquark GPDs are even in skewness
$\eta$
and satisfy the polynomiality constraints.
Moreover, in order to ensure the validity of the factorized description
for the relevant hard processes, GPDs
$F^q$
and
$F^{\bar{q}}$
vanish at
$x=-\eta$
and
$x=\eta$ respectively.

It is convenient to introduce the combinations $F^{(\pm)}$ of GPDs
$F^q$
and
$F^{\bar{q}}$
with  definite charge parity $C=\pm 1$, {\it i.e.}, also with definite symmetry with respect to
$x$:
\be
&&
F^{(+)}(x,\eta,t) = F(x,\eta,t) - \sigma F(-x,\eta,t) = {\rm sign}(x)\left[F^{q}(|x|,\eta,t)+\sigma F^{\bar{q}}(|x|,\eta,t)\right]\,;
\nonumber \\ &&
F^{(-)}(x,\eta,t) = F(x,\eta,t) + \sigma F(-x,\eta,t) = F^{q}(|x|,\eta,t) - \sigma F^{\bar{q}}(|x|,\eta,t)
\,,
\label{Def_singlet_and_nonsinglet_combinations}
\ee
where the signature factor for $H$ and $E$  ($\widetilde H$ and $\widetilde E$) GPDs is $\sigma=1$ ($\sigma=-1$).
We choose skewness
$\eta$
to be positive and, similarly to how it is usually done for quark parton distribution functions (PDFs), consider the
symmetric and the antisymmetric combinations
$F^{(\pm)}(x,\eta,t)$
only for
$x \ge 0$.

For the  case of unpolarized quark GPD
$H$
the combinations
(\ref{Def_singlet_and_nonsinglet_combinations})
appear in the charge even sum of quark and antiquark  (antisymmetric) and the charge odd valence (symmetric)
unpolarized GPDs
$H^{(\pm)}$.
Since the $H$ GPDs will be taken as the main example in the subsequent consideration,
it is worth specifying explicitly our normalization conventions.
\bi
\item In the limit
$\eta \to 0$
the charge even (odd) quark GPD
$H^{(\pm)}$
(\ref{Def_singlet_and_nonsinglet_combinations})
reduces to the sum  (difference) of quark and anti-quark combination of $t$-dependent PDFs:
\be
H^{(\pm)}(x,\eta=0,t) = q(x,t) \pm \overline{q}(x,t)\quad\mbox{with}\quad x\ge 0\,.
\label{Forward_limit_singlet}
\ee
\item  Charge even GPD
$H^{(+)}$
satisfies the polynomiality condition for odd Mellin moments.
Its first Mellin moment is normalized to
the form factors of the quark part of the QCD energy-momentum tensor:
$$
\int_0^1\! dx\, x H^{(+)}(x,\eta,t)= M_2^q(t) + \frac{4}{5} \eta^2 d_1^q(t),
$$
where
$M_2^q(t=0)$
corresponds to the momentum fraction carried by quarks and antiquarks of flavor
$q$
and
$d_1^q(t)$
is the first coefficient of the Gegenbauer expansion of the $D$-term of flavor
$q$, introduced in Ref.~\cite{Polyakov:1999gs}.
\item Charge odd GPD
$H^{(-)}$
satisfies the polynomiality condition for even Mellin moments.
Its zeroth Mellin moment is normalized to
\be
\int_0^1\! dx\,  H^{(-)}(x,\eta,t)= F_1^q(t),
\ee
where
$F_1^q(t)$
stands for the electromagnetic form factor of a particular flavor
$q$.
\ei

For the conformal PW expansion of GPDs we adopt the notations employed in
\cite{Mueller:2005ed}.
The conformal moments of a generic quark GPD
$F$
are formed with respect to the Gegenbauer polynomials
with the index
$\frac{3}{2}$:
$c_n(x,\eta) \sim \eta^n C_n^{\frac{3}{2}} \left( \frac{x}{\eta} \right)$,
where
$n+2$
refers to the conformal spin, labeling the irreducible conformal multiplets of the relevant operators.
Due to the polynomiality property of GPDs and the $T$-invariance, the conformal moments
\be
F_n(\eta,t)=\int_{-1}^1 dx \, c_n(x,\eta) F(x,\eta,t),
\label{Conf_moments}
\ee
are even polynomials in $\eta$ of order $n$ or $n+1$.

The normalization of the conformal basis
\be
c_n(x, \eta)= \eta^n \frac{\Gamma(\frac{3}{2}) \Gamma(1+n)}{2^n \Gamma(\frac{3}{2}+n)} \, C_n^{\frac{3}{2}}\left( \frac{x}{\eta} \right)
\ee
is chosen in such a way that in the forward limit it gives rise to the usual Mellin moments:
\be
\lim_{\eta \to 0} c_n(x,\eta)=x^n.
\ee

It is convenient to introduce the conformal PWs
$p_n(x,\eta)$
including both the integration weight and the support restrictions, expressed by the $\theta$-function,
\begin{eqnarray}
\label{p_n(x,eta)}
p_n(x,\eta) = \eta^{-n-1}  p_n(x/\eta)\,, \quad p_n(x) =
\theta(1-|x|) \frac{2^n \Gamma(\frac{5}{2}+n)}{\Gamma(\frac{3}{2})\Gamma(3+n)} (1-x^2) C^{\frac{3}{2}}_n(-x)\,.
\label{Conformal_PW_basis}
\end{eqnarray}
The conformal PWs
(\ref{Conformal_PW_basis})
are normalized in a way that the following orthogonality relation is valid:
\be
\int_{-1}^1 dx \,p_n(x,\eta) c_m(x,\eta)=(-1)^n \delta_{nm}.
\label{Orthogonality_relation}
\ee
Thus, for integer values of the conformal spin the conformal PWE of a generic quark GPD $F^q$ is formally given by
\begin{eqnarray}
\label{H(x,eta,t)-CPWE}
F(x\ge -\eta,\eta,t) = \sum_{n=0}^\infty (-1)^n p_n(x,\eta) F_n(\eta,t)\,.
\end{eqnarray}
Due to the orthogonality relation
(\ref{Orthogonality_relation}),
the series
(\ref{H(x,eta,t)-CPWE})
obviously reproduces the conformal moments of the GPD
$F^q(x,\eta,t)$.
Since the integral conformal PWs have only $|x|\le \eta$ support, this ill-defined  series can not converge in a common sense.
Loosely spoken one might view the integral conformal PWE (\ref{H(x,eta,t)-CPWE}) as a GPD representation
in the space of singular generalized functions. This series requires rigorous definition in the mathematical sense.

To overcome this issue we convert the integral conformal PWE
(\ref{H(x,eta,t)-CPWE}) the Mellin-Barnes-type integral.
Therefore, we need a  convenient representation
for the conformal PWs for complex valued conformal spin $j+2$. According the Carlson theorem,
this representation uniquely exists if certain assumptions about
the large $j$ asymptotic behavior are satisfied \cite{Carlson:1914}.  The desired representation is  given
by the Schl{\"a}fli integral \cite{Schlaefli}
\begin{eqnarray}
\label{p_n-Schlaefli}
p_j(x,\eta) = -\frac{\Gamma(5/2+j)}{\Gamma(1/2)\Gamma(2+j)} \frac{1}{2i \pi}\oint_{-1}^{1}\!du\, \frac{(u^2-1)^{j+1}}{(x+u\eta)^{j+1}}\,,
\end{eqnarray}
where the integration contour encircles the line segment $u \in [-1;\,1]$.
For integer $j=n$ the Cauchy theorem together with the Rodrigues formula for the Gegenbauer polynomials
\be
\quad (1-x^2)\, C_n^{3/2}(-x) &\!\!\!=\!\!\!& \frac{2+n}{2^{n+1}\, n! } \frac{d^n}{dx^n} (1-x^2)^{n+1}
\ee
allows one immediately to recover the expression for conformal PWs.
Indeed,
\begin{eqnarray}
p_n(x,\eta) &\!\!\!=
\!\!\!& \frac{\Gamma(5/2+n)}{n!\Gamma(1/2)\Gamma(2+n)} \int_{-1}^1\!du\, (1-u^2)^{n+1}\delta^{(n)}(x-u\eta)\,
\end{eqnarray}
yields Eq.\ (\ref{Conformal_PW_basis}).

\subsection{Summing up conformal partial waves with the Mellin-Barnes techniques}
\label{sec:MBI-technique}

In order to illustrate the nature of mathematical subtleties
arising in
(\ref{H(x,eta,t)-CPWE})
it is extremely instructive to consider the limiting case
$\eta=0$.
One may immediately check that the series
(\ref{H(x,eta,t)-CPWE})
provides the formal solution
of the inverse Mellin problem for the relevant
$t$-dependent
parton densities.
Note, that our normalization is chosen in such a manner that for
$\eta=0$ the conformal PWs are simply given by
\be
p_n(x,\eta=0)= \frac{1}{n!}\delta^{(n)}(x).
\ee
Hence, their Mellin moments
$\int_{-1}^1 dx x^m p_n(x,\eta=0)$
give
$(-1)^n \delta_{nm}$.
In particular, for the case of unpolarized quark GPD
$H^q$
one recovers
the Mellin moments of $t$-dependent PDFs
$q(x,t)\equiv H^q(x,\eta=0,t)$.

Obviously, the problem of assigning rigorous meaning to the conformal
PW expansion
(\ref{H(x,eta,t)-CPWE})
is nothing but a special form of the classical moment problem
(see {\it e.g.} \cite{Akhiezer}).
Therefore, the techniques similar to the standard inverse Mellin transform formula, giving rise to
the Mellin-Barnes type integrals along the imaginary axis, turns to be the natural strategy for this issue.
Below we shortly review the approach from Ref.~\cite{Mueller:2005ed} in the form as it is used in global fitting GPD procedures
\cite{Kumericki:2006xx,Kumericki:2009uq,Mueller:2013caa}.

For definiteness, we consider the case of a quark GPD
$F^q$
with the support
$-\eta \le x\le 1$
satisfying the polynomiality condition and vanishing at
$x=-\eta$.
The latter non-perturbative assumption,
which ensures the validity of the relevant factorization theorems,  makes the summation of the series
(\ref{H(x,eta,t)-CPWE})
unique.
Starting from the unphysical case
$\eta>1$
and employing the Sommerfeld-Watson transform one may establish the following Mellin-Barnes integral representation
for GPD
$F^q$:
\begin{eqnarray}
\label{F(x,eta,t)-MB}
F(x\ge -\eta,\eta,t) &\!\!\!=\!\!\!& \frac{1}{2i}\int^{c+i \infty}_{c-i \infty}\!dj\, \frac{ p_{j}(x,\eta)}{\sin(\pi[j+1])} F_{j}(\eta,t)\,.
\end{eqnarray}
This representation
implies the analytic continuation of both the conformal moments
$F_{j}(\eta,t)$
and the conformal PWs
$p_{j}(x,\eta)$
to the complex values of conformal spin.
Special attention should be given to the large-$j$ asymptotic behavior and the analytic properties of the conformal moments.
It is crucial for the validity and uniqueness of the whole procedure ensured by the Carlson theorem (see discussion in
Ref.~\cite{Mueller:2005ed}).

One also needs to specify the position and nature of the rightmost  lying singularities (in $j$) of
$F_{j}$.
This is usually done by taking the pragmatic Regge phenomenology assumptions.
We suppose that the rightmost  lying singularities in
$F_{j}$
are at
$j=\alpha(t)-1$,
where
$\alpha(t)$
stands for the corresponding effective Regge trajectory.
We assume
$\alpha(t=0) <2$
for the $C=+1$ and
$\alpha(t=0) <1$
for the  $C=-1$ combination of GPDs.

Therefore,
the intercept with the real axis $c$
of the integration path
in
(\ref{F(x,eta,t)-MB})
has to be chosen as
$c> {\max} (\alpha-1,-\frac{5}{2})$,
where
$j=-\frac{5}{2}$  is the position of the rightmost pole of the conformal partial waves
({\it cf.} Eqs.~(\ref{ConformalPW:ERBL}) and (\ref{ConformalPW:DGLAP})) %
\footnote{Note that the inclusion of the evolution operator
(or radiative corrections) restricts the lower bound to $c> {\max} (\alpha(0)-1,-1)$.
Also in the case of $H^{(+)}$
the appearance of a perturbative `pomeron' pole requires
$c> {\max} (\alpha(0)-1,0)$.}.
The upper limit for the intercept is
$c <0$.
However, for the antisymmetrized quark GPD combination, {\it e.g.},
$H^{(+)}$,
it can be relaxed to be
$c <1$.
In particular, in the presence of the effective `pomeron' pole in
$H^{(+)}$
the intercept
$c$
is restricted to be
${\max} (\alpha(0)-1,0) < c <1$
with
$\alpha(0) < 2$.

As already stated above, the analytic continuation of the conformal PWs
$p_{j}(x,\eta)$ is provided by the  Schl{\"afli} integral
(\ref{p_n-Schlaefli}).
\bi
\item
For
$|x| \le \eta$
the conformal PWs can be equated with the associated Legendre functions of the first kind,
and expressed {\it  e.g.} in terms of hypergeometric functions
${_2F_1}$:
\begin{eqnarray}
p_{j}(|x| \le \eta,\eta) &\!\!\!=\!\!\!& \frac{2^{j+1}\Gamma(5/2+j)\, \eta^{-j-1}}{\Gamma(1/2)\Gamma(1+j)} (1+x/\eta) {_2F_1}\!\left(-1-j, j+2,2\bigg|\frac{\eta+x}{2\eta}\right)\,.
\label{ConformalPW:ERBL}
\end{eqnarray}
\item For
$x>\eta$
the associated Legendre functions of the first kind have a branch cut. Their discontinuity on
this cut can be expressed in terms of the Legendre functions of the second kind. In terms of hypergeometric functions
it then reads
\begin{eqnarray}
\label{p_j(x,eta)-out}
p_j(x>\eta,\eta)= \frac{\sin(\pi [j+1])}{\pi}\, x^{-j-1} {_2F_1}\!\!\left(\!{(j+1)/2,(j+2)/2 \atop 5/2 + j}\bigg|\frac{\eta^2}{x^2}\right),
\label{ConformalPW:DGLAP}
\end{eqnarray}
where we employed a quadratic transformation for the arguments of the hypergeometric function.
\item Finally, for the region $x\le -\eta$ the conformal PWs are set to zero,
\be
p_{j}(x \le -\eta,\eta)=0\,.
\ee
\ei
It also may be checked that
$p_{j}(x,\eta)$
satisfy the following boundary conditions for the specific values of
$\eta$:
\begin{eqnarray}
\label{p_{j}(x,eta=0)}
p_{j}(x,\eta=0)  &\!\!\!=\!\!\!& x^{-j-1}\, \frac{\sin(\pi [j+1])}{\pi}\,;
 \\
\label{p_{j}(x,eta=x)}
p_{j}(x,\eta=x) &\!\!\!=\!\!\!&  x^{-j-1}\, \frac{2^{j+1} \Gamma(5/2+j)}{\Gamma(3/2)\Gamma(3+j)} \frac{\sin(\pi [j+1])}{\pi}\,.
\end{eqnarray}
Note, that $p_{j}(x,\eta)$ is  continuous in the vicinity of the cross-over line
$x=\eta$.
However, the imaginary part of its first derivative (in $x$)
possesses a discontinuity (while the real part of the first derivative is still continuous).

\section{Dual parametrization}
\label{sec:DualParametrization}
\setcounter{equation}{0}

Another possibility of summing up the conformal PW expansion
(\ref{H(x,eta,t)-CPWE})
is to employ a map of a GPD to  the  forward-like function $\textrm{F}(y,\eta,t)$%
\footnote{The notion ``forward-like function'' is employed to emphasize the simple evolution
properties of
$\textrm{F}(y,\eta,t)$ which are thought to be
described by the usual Dokshitzer-Gribov-Lipatov-Altarelli-Parisi (DGLAP) evolution equation.}
by requiring that the conformal moments
(\ref{Conf_moments})
of the GPD are obtained from  the  Mellin transform in the
auxiliary variable $y$ \cite{Shuvaev:1999fm},
\begin{eqnarray}
F_n(\eta,t) =  \int_{-\left(1+\sqrt{1-\eta^2}\right)/2}^{\left(1+\sqrt{1-\eta^2}\right)/2}\!dy\,y^n\, \textrm{F}(y,\eta,t).
\end{eqnarray}
The integral transformation,
relating the forward-like function $\textrm{F}(y,\eta,t)$ to the initial GPD
$F(x,\eta,t)$, bears the name of the Shuvaev-Noritzsch transform.
For the quark part (\ref{q-part_GPD}) of the GPD $F$ it reads
\be
F(x \ge -\eta ,\eta,t) =  \int_0
^{\left(1+\sqrt{1-\eta^2}\right)/2}\!dy \, K(x,\eta|y) \,\textrm{F}(y,\eta,t)\,,\quad
\ee
The support properties of the forward-like function have been pointed out in
\cite{Noritzsch:2000pr},
but still require to be worked out carefully%
\footnote{
By modeling the conformal moments in terms of the Chebyshev polynomials and employing the inverse Mellin transform,
Noritzsch exemplified
that in order to ensure  polynomiality of the resulting GPDs the forward-like function should possess two branches
with the support
$|y| \le \frac{1+\sqrt{1-\eta^2}}{2}$
expressed by the same function.
However, for a specific GPD model, with the conformal GPD moments defined by a generalization of the Legendre polynomials,
we found out a forward-like function that  possesses the support
$\frac{1-\sqrt{1-\eta^2}}{2} \le y \le \frac{1+\sqrt{1-\eta^2}}{2}\quad  \mbox{rather than}\quad  0\le y \le \frac{1+\sqrt{1-\eta^2}}{2}\,,$
as one would naively expect.}.

The integral kernel $K(x,\eta|y)$ contains  further support restrictions and
its explicit expression was found by means of the dispersive approach
\cite{Shuvaev:1999fm,Noritzsch:2000pr}.
In loose words, it consists
in presenting the conformal PW expansion
(\ref{H(x,eta,t)-CPWE})
as a discontinuity of a certain formal series of generalized functions.
This series is then resummed in the convergency region and afterwards the discontinuity
is taken. Finally, this allows to work out a closed expression for the convolution kernel
$K(x,\eta|y)$
in terms of the elliptic integrals
\cite{Noritzsch:2000pr}.

GPD modeling within the Shuvaev-Noritzsch transform approach can be performed by building up
a model for the forward-like function
$\textrm{F}(y,\eta,t)$
\cite{Noritzsch:2000pr}.
In particular, a simple $\eta$-independent Ansatz with the  forward-like function
given by the corresponding forward PDF (times a $t$-dependent factor) was employed in
\cite{Shuvaev:1999ce}
for approximating GPDs at small values of $\eta$. This model is, strictly spoken, inconsistent
\cite{Noritzsch:2000pr}
and it does not incorporate general GPD properties at small-$\eta$
\cite{Kumericki:2009ji},
as claimed in the literature
\cite{Shuvaev:1999ce}.
Nevertheless, its ``predictive power'' is widely used in the phenomenology of diffractive processes.

In the rest of this Section we introduce the dual parametrization of GPDs
\cite{Polyakov:2002wz}.
It represents a particular receipt for handling the conformal PWE of GPDs
(\ref{H(x,eta,t)-CPWE})
based on a techniques similar to the Shuvaev-Noritzsch transform
in which the conformal moments are further expanded over
a suitable basis of orthogonal
polynomials. As this latter basis one employs the PWs of the cross-channel
${\rm SO}(3)$
rotation group. The dual parametrization techniques explicitly takes care of the polynomiality condition
and, as the result, the support properties of the corresponding forward-like functions are not plagued by the mentioned
above problems and turns to be the same as for PDFs.

\subsection{Basics of the dual parametrization}
\label{sec:BasicsOfDual}

Originally, a series with the structure of a double PWE in the conformal PWs and the
${\rm SO}(3)$ rotation group PWs of the produced
two-hadron system arose for the two-pion generalized distribution amplitude
($2 \pi$ GDA)
\cite{Polyakov:1998ze}.
This object is related to the quark GPD of a pion by the crossing transformation.

\begin{figure}[t]
\centering
\includegraphics[width=7 cm]{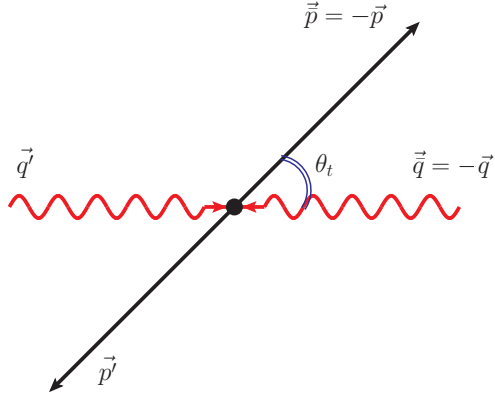}
\caption{\small The definition of the $t$-channel scattering angle in the $t$-channel
center-of-mass reference frame for the reaction
(\ref{DVCS_h}).}
\label{fig:Cos}
\end{figure}

For definiteness, let us consider the DVCS process with a scalar hadron  target
$h$
\be
\gamma^*(q)+h(p) \rightarrow \gamma(q')+h(p')
\label{DVCS_h}
\ee
and its $t$-channel counterpart
\be
\gamma^*(q)+\gamma(-q') \rightarrow  h(p') + \bar{h}(-p).
\label{DVCS_cross}
\ee
For the definition of the $t$-channel scattering angle see Fig.~\ref{fig:Cos}.
Note, that after crossing back to the direct channel, within the DVCS kinematics
($Q^2=-q^2$ and $s\equiv(q+p)^2$ - large;
$\xB \equiv \frac{Q^2}{2 p \cdot q}$ -fixed;
$t\equiv(p'-p)^2 \sim 0$)
the cosine  $\cos \theta_t$ of the $t$-channel scattering angle $\theta_t$ becomes
up to power suppressed corrections
\be
\cos \theta_t \rightarrow -\frac{1}{\eta \beta}+ {\cal O}(1/Q^2),
\ee
where
$\beta=\sqrt{1-\frac{4m^2}{t}}$
stands for the hadron velocity in the $t$-channel center-of-mass frame.
In what follows we switch to massless
($m=0$)
hadrons, so that we could consider
hadron helicities as true quantum numbers and exclude mixing
between the corresponding PW amplitudes.
This implies setting
$\beta=1$
(which means systematically neglecting the threshold corrections
$\sim \sqrt{1-\frac{4m^2}{t}}$).

By exploiting the crossing symmetry, the double PWE, given as a series in the $t$-channel, is analytically
continued to the direct channel and resummed by means of the techniques based on the Shuvaev-Noritzsch transform.
This allows to work out a rigorous expression for the corresponding GPD.
The detailed description of different aspects of the dual parametrization approach can be found in
Refs.~\cite{Polyakov:2007rw,Polyakov:2007rv,Moiseeva:2008qd,SemenovTianShansky:2008mp,Polyakov:2008aa,SemenovTianShansky:2010zv,Pire:2014fwa}.

It is worth now outlining the physical picture constituting the foundation for the dual parametrization of GPDs.
The double PWE for two-hadron GDAs arise naturally within the representation
of the corresponding matrix elements as infinite series of the $t$-channel resonance exchanges of
arbitrary high spin and mass. The selection rules
\cite{Ji:2000id,Diehl:2003ny}
for the quantum numbers of  the $t$-channel resonances make the resulting object satisfy the
requirements following from the discrete symmetries ($C$, $P$ and $T$).
The $t$-channel resonance exchange mechanism can be seen as a development of a purely phenomenological 
model for the matrix elements at a low scale, in the spirit of the vector dominance picture for the 
pion and nucleon electromagnetic form factors \cite{Sakurai:1960ju}. 
On the other hand, at large $N_c$ QCD is believed to become equivalent to a theory of resonances.
This justifies the eventual description of the hadronic matrix elements of
the relevant QCD operators in terms of resonance exchanges.

The double PW expansion representations for two-hadron GDAs are then summed up and analytically continued
to provide the description for the cross-conjugated objects (hadronic GPDs).
Therefore, the term ``dual'' in the appellation of the method alludes
to the natural association with the old idea of duality in hadron-hadron low-energy scattering
\cite{Dolen:1967jr}. For a binary scattering amplitude it can roughly be summarized as the assumption that the infinite sum over only just
the cross-channel Regge exchanges, after appropriate resummation, brings the complete description of the process within certain
kinematical domain
\cite{GreenSchwarzWitten}.
The hope is that a similar mechanism can provide a description of the operator matrix elements occurring in the GPD/GDA definitions.

The dual parametrization approach was initially formulated for the case of a GPD  of a spinless hadron. The generalization
for the spin-$\frac{1}{2}$ target is straightforward. It consists in pointing out the combinations of spin-$\frac{1}{2}$
hadron GPDs suitable for the cross-channel
${\rm SO}(3)$
PW expansion in terms of the (reduced) Wigner rotation $d$-matrices
(see {\it e.g.}~discussion in Sec.~4.2 of Ref.~\cite{Diehl:2003ny}).
For example, for the case of unpolarized nucleon GPDs the $t$-channel helicity conserving
(so-called electric) combination
\be
H^{(E)}(x,\eta,t) \equiv H (x,\eta,t)+ \frac{t}{4m^2} E(x,\eta,t)
\label{Electric_comb_GPDs}
\ee
is to be expanded in
$d_{00}^l(1/\eta) \sim P_l(1/\eta)$,
where $P_l(\chi)$ stand for the Legendre polynomials.
The $t$-channel helicity flip (magnetic) combination
\be
H^{(M)}(x,\eta,t) \equiv H (x,\eta,t)+  E(x,\eta,t)
\label{Magnetic_comb_GPDs}
\ee
is to be expanded in $d_{01}^l(1/\eta) \sim 
P'_l(1/\eta)$.
More details on this subject are presented in Sec.~\ref{sec:MellinBarnesSO(3)}.
Here for simplicity we consider the case of a scalar target GPD%
\footnote{This consideration obviously also applies for the case of the electric combination
 $H^{(E)}(x,\eta,t)$ of unpolarized nucleon GPDs.},
for which the cross-channel
${\rm SO}(3)$-PWE is performed over the usual Legendre polynomials.

Throughout this section we adopt the notations of Refs.~\cite{Polyakov:2002wz,Polyakov:2008aa}.
Our starting point is the double PWE for the charge even
($C=+1$)
and charge odd
($C=-1$)
combinations
(\ref{Def_singlet_and_nonsinglet_combinations})
of the unpolarized quark GPD
$H$:
\begin{eqnarray}
\label{H^{(+)}(x,eta,t)-series}
&&
H^{(+)}(x,\eta,t) = 2\sum_{n=1 \atop {\rm odd}}^\infty \sum_{l=0 \atop {\rm even}}^{n+1}    B_{nl}(t)\, \theta(|\eta|-|x|) \left(\! 1-\frac{x^2}{\eta^2}\right) C_n^{3/2}\left(\frac{x}{\eta}\right) P_l\left(\frac{1}{\eta}\right); \nonumber \\&&
H^{(-)}(x,\eta,t) = 2\sum_{n=0 \atop {\rm even}}^\infty \sum_{l=1 \atop {\rm odd}}^{n+1}    B_{nl}(t)\, \theta(|\eta|-|x|) \left(\! 1-\frac{x^2}{\eta^2}\right) C_n^{3/2}\left(\frac{x}{\eta}\right) P_l\left(\frac{1}{\eta}\right).
\end{eqnarray}
Here, as usual, $n+2$ corresponds to the conformal spin and $l$ refers to the $t$-channel angular momentum
and the generalized form factors
$B_{nl}(t)$
stand for the corresponding double partial wave amplitudes (dPWAs).

Employing the orthogonality of the Gegenbauer polynomials and of the Legendre polynomials,
the dPWAs
$B_{nl}(t)$
can formally be obtained from the ``crossed'' GPD by forming the moments
\begin{eqnarray}
\label{B_{nl}}
B_{nl}(t) = \frac{1}{2}\int_{-1}^1\!d\omega\, \frac{2l+1}{2} P_l(\omega) \int_{-1}^1\!dx\,
\frac{2n+3}{2(n+1)(n+2)}  C_n^{3/2}(x)\,   H^{(+)}\!\left(\frac{x}{\omega},\eta=\frac{1}{\omega},t\right).
\end{eqnarray}
These dPWAs are considered as well defined for any non-negative integer values of $n$ and $l$.

In the spirit of the Shuvaev-Noritzsch transform techniques, we introduce the set of forward-like
functions $Q_{2\nu}(y,t)$ of an auxiliary variable $y$
whose  Mellin transform generates the dPWAs:
\begin{eqnarray}
\label{Q2Bnl}
B_{n,n+1-2\nu}(t)= \int_0^1\! dy\, y^n Q_{2\nu}(y,t)  \quad\mbox{or}\quad
B_{nl}(t)= \int_0^1\! dy\, y^{n} Q_{n+1-l}(y,t).
\end{eqnarray}
The difference of the conformal spin and the $t$-channel angular momentum
$2\nu+1=n+2-l$
with
$\nu\in \{0,1,2,\cdots\}$
is always odd.
A  GPD can now be represented as a series of integral transformations, {\it e.g.},
for the charge even (odd) combination (\ref{H^{(+)}(x,eta,t)-series}):
\begin{eqnarray}
\label{Q22H}
H^{(\pm)}(x,\eta,t)&\!\!\! = \!\!\!& \sum_{\nu=0}^\infty \int_0^1\!dy
\left[ K_{2\nu}(x,\eta|y) \pm K_{2\nu}(-x,\eta|y) \right] y^{2\nu} Q_{2\nu}(y,t)  \,.
\end{eqnarray}
The integral kernels
$K_{2\nu}(x,\eta|y)$
and
$K_{2\nu}(-x,\eta|y)$
defined non-vanishing for
$-\eta \le x \le 1$
and
$-1 \le x \le \eta$
are formally given as the series
\begin{eqnarray}
\label{K_{2nu}(x,eta|y)-sum}
K_{2\nu}(x,\eta|y) &\!\!\! = \!\!\!&
  \sum_{n=-1 }^\infty  \theta(|\eta|-|x|) \left(\!1-\frac{x^2}{\eta^2}\right) C^{3/2}_{n+2\nu}\!\left(\!\frac{x}{\eta}\right)
   y^{n}   P_{n+1}\!\left(\!\frac{1}{\eta}\right)
\end{eqnarray}
with $C_{-1}^{3/2}(z) =0$.
The closed analytic expressions for these kernels were originally calculated by means of Shuvaev's dispersion technique
\cite{Shuvaev:1999ce,Polyakov:2002wz}
and can be expressed in terms of the elliptic integrals.

Note that the convolution over
$y$
in
(\ref{Q22H})
requires some caution.
Therefore, it is worth specifying the small-$y$ behavior of the forward-like functions. The forward-like functions can
be formally reconstructed from GPD moments by the inverse Mellin transform with respect to the complex valued
$n$,
denoted as
$j$,
where
$\nu$
is kept a non-negative integer. It is natural to assume that
for small values of the momentum-type variable
$y$ the forward-like functions inherit the Regge-like behavior of
the Compton form factors at large energy (small $\eta$).
There are two distinct opportunities for this issue.
\bi
\item One may assume that the generalized form factor
$B_{jJ}(t)$
contain the Regge poles in the complex $t$-channel angular momentum $J$-plane,
{\it e.g.}, at
$J\equiv j+1=\alpha(t)$.
Consequently, for all $\nu$ the forward like functions
$y^{2\nu} Q_{2\nu}(y,t)$
have at small
$y$
a
$y^{-\alpha(t)}$
behavior. In this case the convolution integrals in
(\ref{Q22H})
for
$\nu \ge 1$
may require suitable regularization.
\item Contrarily, if one assumes such poles in the complex conformal spin plane at
$j+2\nu+1=\alpha(t)$,
one finds that
$y^{2\nu} Q_{2\nu}(y,t) \sim y^{-\alpha(t)+2\nu}$
and the leading Regge behavior is only present in
$Q_0(y,t)$.
This latter opportunity
(that was, in particular, adopted within the `minimalist' dual parametrization model)
seems, in fact, to be too much restrictive. It results in the skewness ratio
$r^q= \frac{H(x,\eta=x,t=0)}{H(x,\eta=0,t=0)}$
fixed to the so-called conformal value
$r^q_{\rm con}= \frac{2^{\alpha(0)} \Gamma(3/2+\alpha(0))}{\Gamma(3/2)\Gamma(2+\alpha(0))}$
However, this extremely strong model independent prediction
\cite{Shuvaev:1999ce}
contradicts the available
experimental data (see {\it e.g.} Ref.~\cite{Kumericki:2009uq} for the discussion).
\ei

\subsection{Convolution kernels $K_{2\nu}(x,\eta|y)$ from the Schl{\" a}fli integral}
\label{sub_sec_cov_kernels}

In this subsection we present the alternative derivation of the explicit expressions for the
dual parametrization convolution kernels
$K_{2\nu}(x,\eta|y)$
employing  the Schl{\" a}fli integral representation
for the integral conformal PWs
(\ref{p_n(x,eta)}).

It is straightforward to check that
\be
&&
\theta(|\eta|-|x|) \left(\!1-\frac{x^2}{\eta^2}\right) C^{3/2}_{n+2\nu}\!\left(\!\frac{x}{\eta}\right)=
(2+2\nu +n)\frac{(-1)^{n+1}}{2i \pi}\oint_{-1}^{1}\!du\,
\frac{(u^2-1)^{n+1+2\nu}}{(x+u\eta)^{n+1+2\nu}}\, y^{n} \left(\frac{\eta}{2}\right)^{n+1+2\nu}\,.
\nonumber \\ &&
\ee
Plugging this representation into the formal series
(\ref{K_{2nu}(x,eta|y)-sum})
and deliberately interchanging summation and integration orders yields
\begin{eqnarray}
K_{2\nu}(x,\eta|y) = \frac{1}{2i \pi}\oint_{-1}^{1}\!ds\, \sum_{n=-1}^\infty (2+2\nu +n)(-1)^{n+1}
\frac{(s^2-1)^{n+1+2\nu}}{(x-s\eta)^{n+1+2\nu}}\, y^{n} \left(\frac{\eta}{2}\right)^{n+1+2\nu} P_{n+1}\!\left(\!\frac{1}{\eta}\right)
\!.
\nonumber\\
\label{Kernel_from_Shlafli_int}
\end{eqnarray}
In order to match with the notations of
\cite{Polyakov:2002wz},
we rename the integration variable
$u=-s$
in
(\ref{Kernel_from_Shlafli_int})
and introduce the convenient shorthand notation
\be
x_s = \frac{2(x-s\eta)}{(1-s^2)y}\,.
\label{Def_x_s}
\ee
Employing the generating function for the Legendre polynomials,
$$
\sum _{n=-1}^{\infty } \left(\frac{\eta}{x_s} \right)^{n+1} P_{n+1}\!\left(\!\frac{1}{\eta }\right) = \frac{1}{\sqrt{\frac{x_s^2+\eta^2-2 x_s}{x_s^2}}}
$$
we express the convolution kernels
(\ref{Kernel_from_Shlafli_int})
as the following contour integral in the complex $s$-plane:
\begin{eqnarray}
\label{K_{2nu}(x,eta|y)-Schlaefli}
K_{2\nu}(x,\eta|y) &\!\!\!=\!\!\!& \left(\!2+2\nu + y \frac{d}{dy}\right) \frac{1}{2i \pi}\oint_{-1}^{1}\!ds\,
\frac{ \eta^{2 \nu} \left(s^2-1\right)^{2 \nu }}{ 2^{2\nu} (x-s \eta )^{2\nu} y}\\
&&\phantom{ \left(\!2+2\nu + y \frac{d}{dy}\right)\frac{1}{2i \pi}\oint_{-1}^{1}}\times\! \frac{1}{ \sqrt{\frac{y^2 \eta ^2}{4 (x-s \eta )^2}(s-s_1)(s-s_2)(s-s_3)(s-s_4)}}\,.
\nonumber
\end{eqnarray}
Here
$s_i$,
$i=1,\,2,\,3,\,4$
denote the four roots of
$x_s^2+\eta^2-2x_s=0$:
\begin{eqnarray}
\label{s_i}
s_i &\!\!\!=\!\!\!& \frac{1}{y \eta }\left[1-\sqrt{1-\eta ^2} \mp \sqrt{2(1-x y) \left(1-\sqrt{1-\eta ^2}\right)-\left(1-y^2\right) \eta^2}\right] \quad\mbox{for}\quad i = \left\{ {1 \atop 2} \right.\,,\qquad \nonumber
\\
s_i &\!\!\!=\!\!\!& \frac{1}{y \eta }\left[1+\sqrt{1-\eta ^2} \mp \sqrt{2(1-x y) \left(1+\sqrt{1-\eta ^2}\right)-\left(1-y^2\right) \eta ^2}\right]\quad\mbox{for}\quad i = \left\{ {3 \atop 4} \right.\,.\qquad
\end{eqnarray}
To perform the contour integral
(\ref{K_{2nu}(x,eta|y)-Schlaefli})
we need to specify the position
of cuts and poles of the integrand on the real axis in the complex $s$-plane.
For this issue the position and ordering of the roots of the polynomial
$x_s^2+\eta^2-2x_s$
are of major importance.
\bi
\item For
$x\le -\eta$,
assuming
$0\le y\le 1$,
no cuts or poles fall inside the unit circle in the complex $s$-plane.

\item In the central region
$|x| < \eta$
with
$0\le y\le 1$
we get inside the unit circle in the complex $s$-plane the cut
$[s_1,s_3]$.
In addition, for
$\nu>0$  we have in this region
poles at
$s=x/\eta$,
which are of the order
$2\nu-1$.
Note that in this region
$s_3 < x/\eta$.
\item At the cross-over line
$x=\eta$
we recover the following ordering of the roots $s_i$:
$$s_1 =\frac{2-2 \sqrt{1-\eta^2}-\eta y}{\eta y} \le s_2=s_3=  1 \le s_4 =\frac{2+2 \sqrt{1-\eta^2}-\eta y}{\eta y}$$
and, hence, the cut is now in the segment $[s_1,1]$.
\item Finally, in the outer region
$\eta \le x \le 1$
we have inside the unit circle in the complex $s$-plane the cut
$[s_1,s_2]$,
where the
$0 \le y\le 1$
variable is now subject to a lower bound
\begin{eqnarray}
y_0 \le y \quad\mbox{with}\quad
y_0 = \frac{x + \sqrt{x^2-\eta^2}}{1+\sqrt{1-\eta ^2}} \le x \quad\mbox{for}\quad \eta\le x \le 1\,.
\nonumber
\end{eqnarray}
\ei
With the help of this information, we can rewrite the contour integral
(\ref{K_{2nu}(x,eta|y)-Schlaefli})
as the one over the real axes.

For
$\nu=0$
it reads as follows
\begin{subequations}
 \label{K_0(x,eta|y)}
 \begin{eqnarray}
  \label{K_0(xleeta,eta|y)}
 K_0(|x|\le \eta,\eta|y) &\!\!\! = \!\!\!& \frac{1}{y} \frac{d}{dy}\, y
\int_{s_1}^{s_3}\!\frac{ds}{\pi}  \frac{ x_s}{\sqrt{2x_s-x_s^2-\eta^2}}\,;\\
K_0(x\ge \eta,\eta|y) &\!\!\! = \!\!\!&
\frac{1}{y} \frac{d}{dy}\, y  \int_{s_1}^{s_2}\!\frac{ds}{\pi} \frac{\theta(y-y_0) x_s}{\sqrt{2x_s-x_s^2-\eta^2}}\,,
 \label{K_0(xgeeta,eta|y)}
\end{eqnarray}
\end{subequations}
where $x_s$ is defined in
(\ref{Def_x_s})
and
$s_i$
are listed in
(\ref{s_i}).
We note that the integration limits are obtained from the condition
$2x_s-x_s^2-\eta^2\ge 0$.
Therefore, we can write for both regions the same integral
$$
K_0(x\ge -\eta,\eta|y) =
\frac{1}{y} \frac{d}{dy}\, y  \int_{-1}^{1}\!\frac{ds}{\pi} \frac{\theta(2x_s-x_s^2-\eta^2) x_s}{\sqrt{2x_s-x_s^2-\eta^2}}\,,
$$
thus recovering the original result of
\cite{Polyakov:2002wz}.

For
$\nu>0$
the additional pole contribution appear in the central region at
$s=x/\eta$.
It can be easily calculated and can be understood as the piece that regularizes the endpoint singularities of the kernels.
We recover the dual parametrization convolution kernels which we write for positive integer
$\nu$
in a compact form as
\begin{eqnarray}
\label{K_{2nu}(x,eta|y)}
K_{2\nu}(x,\eta|y) &\!\!\! = \!\!\!& \frac{\eta^{2 \nu }}{y^{1+2 \nu }} \frac{d}{dy} y
\int_{-1}^{1}\!ds \Bigg\{
\frac{\theta(2 x_s -x_s^2-\eta^2) x_s^{1-2\nu}}{\pi\sqrt{2 x_s -x_s^2-\eta^2}}
+\frac{y^{2\nu}\delta(x/\eta-s)}{\eta^{2\nu}\, 2^{2\nu}\Gamma(2 \nu)}
\frac{d^{2\nu-1}}{ds^{2\nu-1}} \frac{ x_s (1-s^2)^{2\nu} }{\sqrt{x_s^2+\eta^2-2 x_s }}\Bigg\}.
\nonumber\\
\end{eqnarray}
Employing the generating function for the Legendre polynomials and the Rodrigues formula for the Gegenbauer polynomials one can
show that the subtraction term can be presented as a finite sum
\begin{eqnarray}
\label{K_{2nu}(x,eta|y)-sub}
\frac{y^{2\nu}}{2^{2\nu}\Gamma(2 \nu)}
\frac{d^{2\nu-1}}{ds^{2\nu-1}}\, \frac{x_s (1-s^2)^{2\nu}}{\sqrt{x_s^2+\eta^2-2 x_s}}  \bigg|_{s=\frac{x}{\eta}}\!\!
&\!\!\!\!=\!\!\!\!&
-\theta(|\eta|-|x|)\sum_{l=0}^{2\nu-2} \frac{y^{1+l}}{2+l} \left(\!\!1 - \frac{x^2}{\eta^2}\right)
\!C^{3/2}_{l}\!\!\left(\!\frac{x}{\eta}\!\right)\! P_{2\nu-2-l}\!\!\left(\!\frac{1}{\eta}\!\right)\!.\qquad\quad
\end{eqnarray}
After antisymmetrization/symmetrization in
$x$
it coincides with the known result
\cite{Polyakov:2008aa}
for the charge even (respectively charge odd) combination of GPDs.

A convenient representation for the
convolution kernels
$K_{2\nu}(x,\eta|y)$
can be obtained by separating explicitly the
$J=0$
cross-channel exchange contribution
in the
$|x|< \eta$
region
\be
K_{2\nu}(x,\eta|y) &\!\!\! = \!\!\!& K^{J\neq0}_{2\nu}(x,\eta|y) +
\theta(|\eta|-|x|)  \left(\!1-\frac{x^2}{\eta^2}\right) C^{\frac{3}{2}}_{2\nu-1}\!\!\left(\!\frac{x}{\eta}\!\right) \frac{1}{y},
\ \ \ C^{\frac{3}{2}}_{-1}(x)=0
\label{Kernel_with_J=0_explicit}
\ee
and change of the integration variable
$s \to x_s \equiv z$
in
(\ref{K_{2nu}(x,eta|y)}).
\bi
\item In the central region
$|x|< \eta$
we get
\be
\label{K^{Jneq0}_{2nu}(|x|<eta,eta|y)-dual1}
K^{J\neq0}_{2\nu}(|x|<\eta,\eta|y) &\!\!\! = \!\!\!& \frac{\eta^{2\nu}}{[y^{2\nu}]_+}\int_{z_{2-}}^{z_{2+}}\!\frac{dz}{\pi}\,
\frac{\eta^2-x^2}{\left[(y z -x)^2+\eta ^2-x^2\right]^{\frac{3}{2}}}\,
\frac{z^{1-2 \nu }}{\sqrt{2z-z^2-\eta^2}},
\ee
where
$$
z_{2\pm}= 1\pm\sqrt{1-\eta^2}
$$
are the two roots of
$2z-z^2-\eta^2=0$
and we introduce the ``$+$''-regularization prescription
\cite{GelShi64}
for
$\frac{1}{ y^{2\nu}}$ convolution with a particular test function
$\phi(y)$
by means of a truncated Taylor expansion
\begin{eqnarray}
\quad
\frac{1}{[y^{2\nu}]_+} \phi(y)  = \frac{1}{y^{2\nu}}\left[\phi(y) -\sum_{l=0}^{2\nu-1} \frac{y^l}{l!} \frac{d^l\phi(y)}{dy^l}\Big|_{y=0}\right].
\end{eqnarray}

\item In the outer region $x>\eta$ we get
\be
K^{J\neq0}_{2\nu}(x\ge\eta,\eta|y) &\!\!\! = \!\!\!&
\left[\frac{\eta }{x+\sqrt{x^2-\eta ^2}}\right]^{2 \nu }\frac{ \left(x^2-\eta ^2\right)^{\frac{1}{4}} \sqrt{1+\sqrt{1-\eta ^2}} }{\left(1-\eta ^2\right)^{\frac{1}{4}} \sqrt{x+\sqrt{x^2-\eta ^2}}}\, \delta(y-y_0) \nonumber
\\
&&+\frac{\eta^{2\nu}}{y^{2\nu}}\int_{z_{1+}}^{z_{2+}}\!\frac{dz}{\pi}
\frac{2\theta(y-y_{0}) \left(\eta ^2-x^2\right)}{\left[(y z -x)^2+\eta ^2-x^2\right]^{\frac{3}{2}}\sqrt{z_{2+}-z}}
\!\!\left[\frac{z^{1-2 \nu }}{\sqrt{z-z_{2-}}}
-\frac{z_{1+}^{1-2 \nu }}{\sqrt{z_{1+}-z_{2-}}}\right]
\nonumber\\
{\ }\nonumber\\
&&+\frac{\eta^{2\nu}}{y^{2\nu}}\,\frac{3\sqrt{2}\, \theta(y-y_{0})}{8\sqrt{y}\left(x^2-\eta ^2\right)^{\frac{1}{4}} }\,
\frac{ z_{1+}^{1-2 \nu }}{\sqrt{z_{1+}-z_{2-}}}\;
{_2F_1}\!\!\left(\!{\frac{1}{2},\frac{5}{2}\atop 2}\Big|\frac{z_{1+}-z_{2+}}{z_{1+}-z_{1-}}\!\right),
\label{Kernel_K_outer_region}
\end{eqnarray}
where
$$
z_{1\pm}= \frac{x\pm\sqrt{x^2-\eta ^2}}{y}
$$
are the two roots of
$(y z -x)^2+\eta ^2-x^2=0$.
\ei

Note that in the central region
$z_{1\pm}$ become
complex-valued and, hence, we get only integrable singularities in
(\ref{K^{Jneq0}_{2nu}(|x|<eta,eta|y)-dual1}).
In the outer region we performed the regularization
of the endpoint singularity at
$z=z_{1+}$
in such a manner that the integral in the second line
of
(\ref{Kernel_K_outer_region})
converges and can be numerically evaluated in a straightforward way.

The pole contribution, arising in the kernels
(\ref{K_{2nu}(x,eta|y)})
for
$\nu>0$,
provides a regularization of the convolution integral
(\ref{Q22H})
for the central region in the limit
$y\to 0$.
However, once we assume that the small-$y$
asymptotic behavior of forward-like functions
is determined by the leading Regge singularity in the complex-$J$ plane
$y^{2 \nu} Q_{2 \nu}(y,t) \sim y^{-\alpha(t)}$
(see discussion in Sec.~\ref{sec:BasicsOfDual}),
a
$1/y$
divergence still occurs in the convolution integral
(\ref{Q22H})
for the case of the charge even quark GPD combination
$H^{(+)}$.
Indeed, one may check that this
divergence is manifest from the very beginning for
$J=0$
dPWAs
\be
B_{2\nu-1 \,0}(t)= \int_0^1 \frac{dy}{y} y^{2 \nu}  Q_{2 \nu}(y,t),
\label{Pot_divergent_FFs}
\ee
and  require pointing out a suitable regularization.
This ambiguity is closely related to the problem of the
$J=0$ cross-channel exchange contribution into the
$D$-term part of GPDs. Depending on the representation, this problem
can be formulated in various languages, and, therefore, it is
important to distinguish between the mathematical aspects and the physical contents.

From the mathematical point of view, the problem of providing
a regularization to (\ref{Pot_divergent_FFs})
corresponds to assigning meaning for a convolution of a generalized function
with a power-like singularity at
$y=0$
with a test functions belonging to a suitable class
(see {\it e.g.} discussion in \cite{GelShi64}). There exist an infinite
number of such regularizations,  which differ at most by a constant value.
Choosing some particular regularization from this infinite set requires attracting
some external physical principle.

In fact, a similar problem of assigning meaning to the inverse Mellin moments of the scattering amplitude was encountered in the
$S$-matrix theory
\cite{Khuri:1963zza}.
Following that experience, a tempting possibility to deal with (\ref{Pot_divergent_FFs})
is to adopt a suitable analyticity assumption
(so-called ``analyticity of the second kind'', see {\it e.g.} Chapter~I of \cite{Alfaro_red_book})
allowing to fix the values of the divergent dPWAs
by the analytic continuation of dPWAs
$B_{nJ}(t)$
in the angular momentum
$J$ to $J=0$.
This corresponds to the use of the analytic regularization prescription
\cite{GelShi64} in (\ref{Pot_divergent_FFs}):
\be
B_{2\nu-1 \,0}(t)= \int_{(0)}^1 \frac{dy}{y} y^{2 \nu}  Q_{2 \nu}(y,t),
\label{Pot_divergent_FFs-AC}
\ee
where
$(0)$
denotes the regularization procedure. To give a concrete example, let us assume that  in the
$y\to 0$
limit
$ y^{2 \nu}  Q_{2 \nu}(y,t)$
behaves as
$y^{-\alpha(t)}  Q^{\rm res}_{2 \nu}(t)$
and
$y^{2 \nu}  Q_{2 \nu}(y,t)- y^{-\alpha(t)}  Q^{\rm res}_{2 \nu}(t)$
vanishes in this limit.
Hence, the integral can be rewritten as
 \be
&&
B_{2\nu-1 \,0}(t)= \int_{(0)}^1 \frac{dy}{y} y^{-\alpha(t)}  Q^{\rm res}_{2 \nu}(t) +  \int_{0}^1 \frac{dy}{y} \left[  y^{2 \nu}  Q_{2 \nu}(y,t)- y^{-\alpha(t)}  Q^{\rm res}_{2 \nu}(t)\right] \nonumber \\  &&
=-\frac{Q^{\rm res}_{2 \nu}(t)}{\alpha(t)} +  \int_{0}^1 \frac{dy}{y} \left[  y^{2 \nu}  Q_{2 \nu}(y,t)- y^{-\alpha(t)}  Q^{\rm res}_{2 \nu}(t)\right],
\label{Pot_divergent_FFs-AC1}
\ee
where the last integral can now be evaluated straightforwardly.

Despite looking appealing from the theory point of view, the suggested analyticity assumption
is not necessarily fulfilled. In particular, it can be violated by admitting the so-called  $J=0$ fixed pole (f.p.)
contribution manifest as the Kronecker-delta singularities for dPWAs
\be
B_{nJ}(t) \to B_{nJ}(t) + \delta_{0J} B^{\rm f.p.}_{n0}(t)\,,
\label{Fixed_pole_to_Bnl}
\ee
which can not be revealed by the analytical continuation in
$J$.
The inverse moment of
$y^{2 \nu}  Q_{2 \nu}(y,t)$
in
(\ref{Pot_divergent_FFs})
is then defined as
\begin{eqnarray}
B_{2\nu-1 \, 0}(t)= {\rm Reg}\int_{0}^1 \frac{dy}{y} y^{2 \nu}  Q_{2 \nu}(y,t)
= \int_{(0)}^1 \frac{dy}{y} y^{2 \nu}  Q_{2 \nu}(y,t) + B^{\rm f.p.}_{2\nu-1,0}(t).
\label{Pot_correct_FFs}
\end{eqnarray}
Note that equivalently the fixed pole contribution
(\ref{Fixed_pole_to_Bnl})
can be formally included as an ``invisible term''
in the forward-like functions%
\footnote{We define $\int_0^1 dx \delta(x)= \frac{1}{2} \int_{-1}^1 dx \delta(x)= \frac{1}{2}$.}
$$ y^{2\nu} Q_\nu(y) \to  y^{2\nu} Q_\nu(y) + 2 y \delta(y) \,  B^{\rm f.p.}_{2\nu-1 \,0}(t).$$

A particular example of a GPD model with a non-zero
$J=0$
fixed pole contribution is provided by the calculation
\cite{SemenovTianShansky:2008mp}
of pion GPDs in the non-local chiral quark model \cite{Praszalowicz:2003pr}.
In this model the analyticity assumption (\ref{Pot_divergent_FFs-AC}) is not valid due to a $J=0$ fixed pole contribution,
which has to be added to make GPD satisfy the soft pion theorem
\cite{Pobylitsa:2001cz}
fixing pion GPDs in the limit
$\eta \to 1$.

Thus, the problem of appropriate choice of regularization for
(\ref{Pot_divergent_FFs})
can also be formulated on a somewhat old-fashioned $S$-matrix theory language as ``Is there a
$J=0$
fixed pole  contribution into dPWAs?''. The use of the analytic regularization implies the absence of a
$J=0$
fixed pole contribution into dPWAs, whereas employing of an alternative regularization
(differing by a constant) corresponds to a non-zero
$J=0$
fixed pole  contribution. This issue is discussed in more details in
Sec.~\ref{sec: SO(3)PW and fixed pole}
in the related context of
$J=0$
fixed pole  contribution into the
$D$-form factor.

\section{SO(3) partial wave expansion with Mellin-Barnes integral}
\label{sec:MellinBarnesSO(3)}
\setcounter{equation}{0}

The cross-channel
${\rm SO}(3)$-PW
expansion of the conformal PWs, discussed in
Sec.~\ref{sec:BasicsOfDual}
in the dual parametrization framework,
has been also implemented  within the approach based on the Mellin-Barnes
integral representation
(\ref{F(x,eta,t)-MB}),
see
\cite{Mueller:2005ed,Kumericki:2009uq}.
Thereby, the authors
expanded GPD conformal moments in terms of the reduced Wigner $d$-functions.
In this section we establish the link between the  two approaches
and show that
the Mellin-Barnes
integral representation with the cross-channel
${\rm SO}(3)$-PW
expansion
and the dual parametrization of GPDs
are nothing but the two dialects of the same language.

For simplicity, we again consider the case of a quark GPD
$H$ of a spin-$0$ hadron%
\footnote{ This consideration also literally applies for the electric combination
(\ref{Electric_comb_GPDs})
of unpolarized nucleon GPDs and can be trivially extended for the case of the magnetic combination
(\ref{Magnetic_comb_GPDs}).}.
The cross-channel
${\rm SO}(3)$-PW
expansion  of the corresponding conformal moments (\ref{Conf_moments}) reads
\begin{eqnarray}
\label{H_n(eta,t)-SO(3)}
&&
H_n(\eta,t)= \sum_{\nu=0}^{(n+1)/2} \eta^{2\nu} H_{n,n+1-2\nu}(t) \hat{d}^{n+1-2\nu}(\eta), \ \ \ {\rm for \ \ odd} \ \ n;
\nonumber \\ &&
H_n(\eta,t)= \sum_{\nu=0}^{n/2} \eta^{2\nu} H_{n,n+1-2\nu}(t) \hat{d}^{n+1-2\nu}(\eta), \ \ \ {\rm for \ \ even} \ \ n,
\end{eqnarray}
where the ${\rm SO}(3)$-PWs
$\hat{d}^l_{00}$
are expressed by the reduced Wigner
$d^l_{00}$-functions that for a scalar hadron are the Legendre polynomials
\be
\hat{d}^l_{00}(\eta) =
\frac{\Gamma\!\left(\frac{1}{2}\right) \Gamma(1+J)}{2^J \Gamma\!\left(\frac{1}{2}+J\right)} \eta^l  P_l\!\left(\!\frac{1}{\eta}\!\right)\,.
\label{Def_Reduced_d_function}
\ee
The normalization of the ${\rm SO}(3)$-PWs in
(\ref{Def_Reduced_d_function})
is chosen in a way that
$\hat{d}^J_{00}(\eta=0)=1$.

Plugging the expansion
(\ref{H_n(eta,t)-SO(3)})
into the Mellin-Barnes integral representation provides us
the conformal and cross-channel ${\rm SO}(3)$ partial wave expansion of the
quark part
(\ref{q-part_GPD})
of the GPD $H$
\begin{eqnarray}
\label{nu-expansion}
H(x\ge -\eta,\eta,t) &\!\!\!=\!\!\!& \sum_{\nu=0}^\infty \frac{1}{2i}\int^{c+2\nu+i \infty}_{c+2\nu-i \infty}\!dj\,
\frac{ p_{j}(x,\eta)}{\sin(\pi[j+1])}\, H_{j,j+1-2\nu}(t)\, \eta^{2\nu} \hat{d}^{j+1-2\nu}_{00}(\eta)
\\
&&\!\!\!\!\!
-
\sum_{\nu=1}^\infty
\eta^{2\nu}\,  p_{2\nu-1}(x,\eta) H_{2\nu-1,0}(t)\,.
\nonumber
\end{eqnarray}
Here the intercept $c$ of the Mellin Barnes integral contour is to be taken as specified in Sec.~\ref{sec:MBI-technique}
and the integration path is shifted to the right by
$2\nu$
units for partial waves with
$\nu >0$
in order to ensure the polynomiality condition.
Note that the second term in the r.h.s.~of
(\ref{nu-expansion}),
corresponding to
the eventual
$J=0$
exchange contribution,
was omitted in the
small-$\xB$
consideration of
Ref.~\cite{Kumericki:2009uq}%
\footnote{
The $J=0$ exchange contribution that plays a minor role at small-$\xB$
was implicitly taken into account in the KM hybrid model trough a subtraction constant, see fixed pole discussion in
Sec.~\ref{sub_sec_cov_kernels} and below in Sec.~\ref{sec: SO(3)PW and fixed pole}.
\label{note:subtraction_term}}.
Now renaming the integration variable
$j\to j+2\nu$
yields the following double PWE
for the quark part of GPD
$H$:
\begin{eqnarray}
\label{F(x,eta,t)-SO3-MB}
&&
H(x\ge -\eta,\eta,t)  \nonumber     =\sum_{\nu=0}^\infty  \frac{1}{2i}\int^{c+i \infty}_{c-i \infty}\!dj\,
\frac{\eta^{2\nu} p_{j+2\nu}(x,\eta)}{\sin(\pi[j+1])}\, H_{j+2\nu,j+1}(t)\, \hat{d}^{j+1}_{00}(\eta) \\ &&
-
\sum_{\nu=1}^\infty
\eta^{2\nu}\,  p_{2\nu-1}(x,\eta) H_{2\nu-1,0}(t)\,.
\end{eqnarray}
Therefore, assuming
$x\ge 0$,
we obtain the following representation for the charge even  GPD combination
(\ref{H^{(+)}(x,eta,t)-series}):
\begin{eqnarray}
\label{H^+(x,eta,t)-SO3-MB}
H^{(+)}(x,\eta,t) &\!\!\!=\!\!\!&\sum_{\nu=0}^\infty  \frac{1}{2i}\int^{c+i \infty}_{c-i \infty}\!dj\,
\frac{\eta^{2\nu}\left[ p_{j+2\nu}(x,\eta)-p_{j+2\nu}(-x,\eta)\right]}{\sin(\pi[j+1])}\, H^{}_{j+2\nu,j+1}(t)\, \hat{d}^{j+1}_{00}(\eta)
\nonumber\\
&&\!\!\!\!\!-
\sum_{\nu=1}^\infty
\eta^{2\nu}\,  2p_{2\nu-1}(x,\eta) H^{}_{2\nu-1,0}(t)\,.
\end{eqnarray}

It is now straightforward to establish the formal link between the dPWA
$H_{nl}(t)$,
occurring in
(\ref{H_n(eta,t)-SO(3)}), and the set of the forward-like functions
$Q_{2\nu}(y,t)$,
introduced in the context of the dual parametrization.
Indeed, the dPWA
$H_{nl}(t)$
can be put in correspondence with the generalized form factors
$B_{nl}(t)$
given by the Mellin moments (\ref{Q2Bnl}) of the forward-like functions:
\begin{eqnarray}
\label{HjJ2B}
 H^{}_{n,n+1-2\nu}(t) = \frac{\Gamma(3+n)\Gamma\!\left(\frac{3}{2}+n-2\nu\right)}{2^{2\nu} \Gamma\!\left(\frac{5}{2}+n\right)\Gamma(2+n-2\nu)} B_{n,n+1-2\nu}(t).
\end{eqnarray}
The inversion of the corresponding Mellin transform allows to reconstruct the forward-like functions from the dPWAs,
\begin{eqnarray}
\label{HjJ2Q}
y^{2\nu} Q_{2\nu}(y,t) = \frac{1}{2\pi i} \int_{c-i \infty}^{c+i \infty}\!dj\, y^{-j-1}\,\frac{2^{2\nu} \Gamma(5/2+j+2\nu)\Gamma(2+j)} {\Gamma(3+j+2\nu)\Gamma(3/2+j)} H_{j+2\nu,j+1}(t)\,.
\end{eqnarray}

The next step is to plug the expression
(\ref{HjJ2B})
for the partial wave amplitudes
$H_{nl}(t)$
through the Mellin moments
(\ref{HjJ2B})
into the corresponding Mellin-Barnes integral representation.
By comparing the result with
Eqs.~(\ref{Q22H}),
(\ref{Kernel_with_J=0_explicit}),
one can read off the following Mellin-Barnes representation for
the dual parametrization convolution kernels
$K_{2\nu}(x,\eta|y)$:
\begin{subequations}
\label{K_{2nu}(x,eta|y)-MB}
\begin{eqnarray}
\label{K_{2nu}(x,eta|y)-MBa}
K_{2\nu}(x,\eta|y) &\!\!\!=\!\!\!& K^{J\neq0}_{2\nu}(x,\eta|y) - \eta^{2\nu}\,  p_{2\nu-1}(x,\eta)
\frac{\Gamma\!\left(\frac{1}{2}\right)\Gamma(2+2\nu)}{2^{2\nu} \Gamma\!\left(\frac{3}{2}+2\nu\right)}\, \frac{1}{y}
\phantom{\Bigg]};
\\
\label{K_{2nu}(x,eta|y)-MBb}
K^{J\neq0}_{2\nu}(x,\eta|y) &\!\!\!=\!\!\!&  \frac{1}{2i}\int^{c+i \infty}_{c-i \infty}\!dj\,  \eta^{2\nu}\,
\frac{p_{j+2\nu}(x,\eta)}{\sin(\pi[1+j])}\, \frac{\Gamma(3+j+2\nu)\Gamma\!\left(\frac{3}{2}+j\right)}{2^{2\nu} \Gamma\!\left(\frac{5}{2}+j+2\nu\right)\Gamma(2+j)}\; y^{j}\,  \hat{d}^{j+1}_{00}(\eta)\,.
\end{eqnarray}
\end{subequations}
Here again the contour intercept $c$ is chosen as explained in Sec.~\ref{sec:MBI-technique}.

Let us show that the Mellin-Barnes representation
(\ref{K_{2nu}(x,eta|y)-MB})
of the dual parametrization convolution kernels
$K_{2\nu}(x,\eta|y)$
in fact coincides with the integral representations
(\ref{K_{2nu}(x,eta|y)}).
The conformal partial waves can be expressed through the Schl\"afli integral
(\ref{p_n-Schlaefli}).
In terms of the
$s=-u$
variable they read as
\begin{eqnarray}
\frac{(\eta/2)^{2\nu}\Gamma(3+j+2\nu)}{\Gamma\!\left(\frac{5}{2}+j+2\nu\right)}\frac{p_{j+2\nu}(x,\eta)}{\sin(\pi [j+1])} &\!\!\!=\!\!\!&
\frac{(\eta/2)^{2\nu}(2+2\nu+j)}{\pi \Gamma\!\left(\frac{1}{2}\right)} \int_{-1}^{(s_\eta)}\!du\,\frac{\left(1-s^2\right)^{j+1+2\nu}}{(x-s \eta)^{j+1+2\nu}}\,,
\label{Integral_rep_for_conf_PWs}
\end{eqnarray}
where
\be
s_\eta=
\begin{cases}
1, \ \ {\rm for} \ \  x \ge \eta \\
x/\eta  , \ \ {\rm for} \ \  |x| < \eta
\end{cases}.
\ee
The use of the analytic regularization prescription, denoted in the upper integration limit as
$(s_\eta)$, is implied for the pole singularity at $s=\frac{x}{\eta}$.
In the outer region these integrals are convergent and the upper integration limit is given by $1$.
Furthermore, the SO(3)-PWs $\hat{d}^{j+1}_{00}(\eta)$, defined in Eq.~(\ref{Def_Reduced_d_function}),
might be presented as the integral
\begin{eqnarray}
\frac{\Gamma\!\left(\frac{3}{2}+j\right)}{\Gamma(2+j)}\hat{d}^{j+1}_{00}(\eta) = \frac{\Gamma\!\left(\frac{1}{2}\right) \eta^{j+1} }{2^{j+1}} P_{j+1}\left(\frac{1}{\eta} \right)\,
= \frac{\Gamma\!\left(\frac{1}{2}\!\right)}{\pi} \int_{-1}^{1}\!\frac{dv}{\sqrt{1-v^2}}\,\left(\!\frac{1+v\sqrt{1-\eta ^2} }{2}\right)^{j+1}\,.
\label{Integral_rep_for_dhat}
\end{eqnarray}

Plugging the two integral representations
(\ref{Integral_rep_for_conf_PWs}),
(\ref{Integral_rep_for_dhat})
into the Mellin-Barnes representation
(\ref{K_{2nu}(x,eta|y)})
finally allows us to perform the inverse Mellin transform, which yields%
\footnote{We also could set the upper limit of the $s$-integral to
$1$
(certainly, implying the analytical regularization prescription for the pole in the central region),
since the integration region is actually  properly taken into account by the support of the function that arises from the $v$-integration.}
\begin{eqnarray}
K^{J\neq0}_{2\nu}(x,\eta|y)&\!\!\! =\!\!\!& \frac{\eta^{2\nu}}{y^{2\nu+1}} \frac{d}{dy}\, y \, \frac{1}{\pi} \int_{-1}^{(s_\eta)}\!ds\, x_s^{1-2v} \int_{-1}^1\!\frac{dv}{\sqrt{1-v^2}}\;
 \delta(x_s - 1-v \sqrt{1-\eta^2})\,,
\end{eqnarray}
where
$x_s$
was defined in Eq.~(\ref{Def_x_s}).
Interchanging the integration order and performing the $v$-integration first,
$$
\int_{-1}^1\!\frac{dv}{\sqrt{1-v^2}}\; \delta(x_s - 1-v \sqrt{1-\eta^2}) =
\frac{\theta\left(1-\eta^2-(1-x_s)^2\right)}{\sqrt{1-\eta^2}\;\sqrt{1-\frac{(1-x_s)^2}{1-\eta^2}}} = \frac{\theta\left(2x_s-x_s^2-\eta^2\right)}{\sqrt{2x_s-x_s^2-\eta^2}}\,,
$$
we recover  the familiar dual parametrization result for the convolution kernels
(\ref{K_{2nu}(x,eta|y)}),
including the subtraction term
(\ref{K_{2nu}(x,eta|y)-sub}).
Consequently, we also have established that the Mellin moments of the convolution kernels provide the conformal partial waves
\begin{eqnarray}
\int_{(0)}^\infty\!dy\, y^{-j-1} K^{J\neq0}_{2\nu}(x,\eta|y)  &\!\!\!=\!\!\!&
\frac{\pi p_{j+2\nu}(x,\eta)}{\sin(\pi[1+j])}\,  \frac{\Gamma(3+j+2\nu)\Gamma\!\left(\frac{3}{2}+j\right)}{2^{2\nu} \Gamma\!\left(\frac{5}{2}+j+2\nu\right)\Gamma(2+j)}\, \hat{d}^{j+1}_{00}(\eta)\,,
\end{eqnarray}
where $(0)$ stands for the analytic regularization prescription
(see
Eq.~(\ref{Pot_divergent_FFs-AC1})).

\section{Special limiting cases}
\label{sec:SpecialCases}
\setcounter{equation}{0}

For illustrative purposes it is extremely instructive to consider
some special limiting cases, in which the dual parametrization convolution kernels
$K_{2 \nu}(x,\eta|y)$
can be straightforwardly derived from the Mellin-Barnes representation
(\ref{K_{2nu}(x,eta|y)-MB})
and compared to the known result
(\ref{K_{2nu}(x,eta|y)})
in the dual parametrization framework. For definiteness,
throughout this section we consider the case of charge even
spin-$0$ target quark GPD
$H^{(+)}$. The generalization for spin-$\frac{1}{2}$ target case is straightforward.

\subsection{$t$-dependent parton densities ($\eta=0$)}
\label{ssec:fwd}
As a first example, we consider
the limit
$\eta=0$, in which GPD $H^{(+)}$ is reduced to the corresponding
$t$-dependent PDF
(\ref{Forward_limit_singlet}).
For
$\eta=0$
only the
$\nu=0$
kernel
(\ref{K_0(x,eta|y)})
in the outer region
\begin{eqnarray}
K_{0}(x,\eta=0|y)=\left[\delta(y-x)+\theta(y-x)\,\frac{ x^{1/2}}{2y^{3/2}}\right]\,,
\label{Forward_limit_kernel}
\end{eqnarray}
is relevant.
The GPD is therefore expressed as
\be
\label{Q_02H(x,0,t)}
H^{(+)}(x,0,t)&\!\!\!=\!\!\!& Q_0(x,t) + \frac{\sqrt{x}}{2}\int_x^1\!\frac{dy}{y}\, \frac{Q_0(y,t)}{\sqrt{y}}\,.
\ee

In  the Mellin-Barnes integral representation the conformal PWs reduce  to the inverse Mellin
transform integral kernel. Therefore, the
$t$-dependent parton density is given by
\begin{eqnarray}
\label{Q_02H(x,0,t)_MB}
H^{(+)}(x,0,t)
=\frac{1}{2\pi\, i}\int^{c+i \infty}_{c-i \infty}\!dj\,  x^{-j-1} H_{j,j+1}(t)\,.
\nonumber
\end{eqnarray}
Employing
(\ref{HjJ2B})
we express the corresponding dPWA as
\be
H^{(+)}_{j,j+1}(t) = \frac{2(2+j)}{3+2j} \int_0^1\! dy\, y^{j} Q_{0}(y,t)
\ee
and immediately recover the familiar dual parametrization result for the kernel
(\ref{Forward_limit_kernel})
since
\be
 \frac{1}{2\pi i}\int^{c+i \infty}_{c-i \infty}\!dj\,   (x/y)^{-j-1} \frac{2(2+j)}{(3+2j)y} = \delta(x-y)
 + \frac{\theta(1-x/y)}{y}\, \frac{x^{1/2}}{2y^{1/2}}\,.
\ee

\subsection{ GPD on the cross-over line  ($x=\eta$)}
\label{sec-cases_cross-over}

The GPD behavior on the cross-over line
$x=\eta$
is of special importance since it corresponds to the imaginary part of the
leading order elementary amplitude of hard exclusive reactions.
Therefore, GPDs on the cross-over line have direct relation to the observable quantities
({\it e.g.} the Compton form factors).

The convolution kernels for the GPD on the cross-over line can be straightforwardly obtained from the general expression
within the dual parametrization framework
(\ref{K_{2nu}(x,eta|y)}).
They turn to be independent of the index
$\nu$
and read as following:
\begin{eqnarray}
\label{K_{2nu}(x,x|y)}
K_{2\nu}(x,\eta=x|y)=\theta\!\left(\!y-\frac{x}{1+\sqrt{1-x^2}}\right)\frac{1}{y}\frac{2}{\pi\, \sqrt{\frac{2y}{x}-1-y^2}}\,.
\end{eqnarray}
This yields the familiar integral transform for the GPD at the cross-over line
\be
\label{H+(x,x)-dual}
H^{(+)}(x,x,t) &\!\!\!=\!\!\!&   \frac{1}{\pi}\int_{\frac{x}{1+\sqrt{1-x^2}}}^1\!\frac{dy}{y}\,
\frac{2}{\sqrt{\frac{2y}{x}-1-y^2}} \, N(y,t) \,,
\ee
where
$N(y,t)$
stands for the so-called GPD quintessence function
\cite{Polyakov:2007rv}
\be
N(y,t)=\sum_{\nu=0}^\infty y^{2\nu} Q_{2\nu}(y,t)\,.
\label{GPD_quintessence}
\ee

Now, by considering the Mellin-Barnes integral representation
(\ref{F(x,eta,t)-SO3-MB})
for
$\eta=x$,
we get
\begin{eqnarray}
H^{(+)}(x,x,t)
\label{H+(x,x)-MB}
 &\!\!\!=\!\!\!&   \frac{1}{2\pi i}\int^{c+i \infty}_{c-i \infty}\!dj\, x^{-j-1} \, \hat{d}^{j+1}_{00}(x)
\sum_{\nu=0}^\infty \frac{2^{j+1+2\nu}\Gamma\!\left(\frac{5}{2}+j+2\nu\right)}{\Gamma\!\left(\frac{3}{2}\right)\Gamma(3+j+2\nu)}  H^{}_{j+2\nu,j+1}(t)\,.
\end{eqnarray}
Expressing the relevant dPWAs through
the Mellin moments of the forward-like functions
with the help of Eq.~(\ref{HjJ2B})
and employing the explicit expression for the
${\rm SO}(3)$-PWs
(\ref{Def_Reduced_d_function})
we find out
the Mellin representation of the integral kernel
\begin{eqnarray}
\label{nu-expansion_cross-over}
K_{2\nu}(x,\eta=x|y)= \frac{1}{2\pi i}\int^{c+i \infty}_{c-i \infty}\!dj\, y^{j}\, 2P_{j+1}(1/x)
\qquad
\end{eqnarray}
and recover
Eq.~(\ref{K_{2nu}(x,x|y)}).
We also conclude that the Mellin transform of the dual parametrization convolution kernels
on the cross-over line gives the Legendre polynomials
\be
\int_{0}^{1} dy\, y^{-j} K_{2\nu}(x,\eta=x|y) = 2P_j(1/x)\,.
\ee

\subsection{Meson-like GPD ($\eta=1$)}
\label{ssec:eta_1}
Another limiting case, in which the convolution kernels
(\ref{Kernel_with_J=0_explicit})
greatly simplify, is the limit
$\eta \to 1$.
The properties of GPDs in this limit are similar to those of meson distribution
amplitudes.
With the help of a straightforward calculation
one can check that the dual parametrization convolution kernels reduce to
\begin{eqnarray}
\label{K_{2nu}(x,1|y)}
K_{2\nu}(x,\eta=1|y)=y^{-2 \nu } \left[\frac{1-x^2}{\left(1-2 x y+y^2\right)^{3/2}} -\sum_{l=0}^{2\nu-2} y^{l}
\left(1 -x^2\right)\!C^{3/2}_{l}(x)\right].
\end{eqnarray}
In fact, this is nothing but the subtracted generating function for the Gegenbauer polynomials.

The same expression can be derived from the Mellin-Barnes integral representation
(\ref{K_{2nu}(x,eta|y)-MB}).
By setting
$\eta=1$
with
$\hat{d}^{j+1}_{00}(1)= \Gamma(1/2) \Gamma(2+j)/2^{j+1}  \Gamma\left(3/2+j\right)$,
we obtain the Mellin-Barnes integral for the meson-like GPD,
\begin{eqnarray}
\label{K_{2nu}(x,eta=1|y)-MB}
K_{2\nu}(x,\eta=1|y) &\!\!\!=\!\!\!&  \frac{1}{2i}\int^{c+i \infty}_{c-i \infty}\!dj\,
\frac{p_{j+2\nu}(x)}{\sin(\pi[1+j])}\,  \frac{\Gamma\!\left(\frac{1}{2}\right)\Gamma(3+j+2\nu)}{2^{j+1+2\nu} \Gamma\!\left(\frac{5}{2}+j+2\nu\right)}\; y^{j}
\end{eqnarray}
where
$p_j(x) =p_j(x,\eta=1)$.
Since for large
$j$
$$
\frac{p_{j+2\nu}(x\pm i\epsilon)}{2^{j+1+2\nu} \sin(\pi[j+1])} \sim \frac{e^{(j+2 \nu ) {\rm arccosh}(-x\pm i\epsilon)} \pm i e^{-(j+3+2 \nu ){\rm arccosh}(-x\pm i\epsilon))}}{\sin(\pi[j+1])}
$$
we may change the integration path so that the r.h.s.~of the real axis is encircled.
Picking up the residues for non-negative integer
$j$
provides
the convergent series
(\ref{nu-expansion})
for
$\eta=1$.
The kernel then reads
\begin{eqnarray}
K_{2\nu}(x,\eta=1|y) = \sum_{n=0}^\infty  \left(1-x^2\right) C^{3/2}_{n+2\nu}(x)\, y^{n}
\quad \mbox{with}\quad  C^{3/2}_{-1}(x)=0\,,
\end{eqnarray}
which corresponds to the Gegenbauer polynomial generating function
(\ref{K_{2nu}(x,1|y)}).

\subsection{ $D$-term extraction ($\eta\to \infty$)}
\label{ssec:D_term}

Finally, we would like to discuss GPDs in the in the unphysical region by taking the
`low energy' limit $\eta \to \infty$ which allows us to extract the $D$-term.
Rescaling of $x$ with $\eta$, {\it i.e.}, $x\to \eta x$,
and taking the $\eta\to\infty$ limit yields
\begin{eqnarray}
D(x,\eta,t)= \theta(|x|\le |\eta|) d(x/|\eta|,t)\,,\quad
d(x,t)= \lim_{\eta\to \infty} H(x\eta,\eta,t)\,.
\label{Def_D_term}
\end{eqnarray}
This procedure implies that the poles in the complex $s$-plane
in the integrand of
Eq.~({\ref{K_{2nu}(x,eta|y)})
are getting imaginary and that the corresponding Mellin-Barnes integral is only defined
for negative $x$-values.
Formally, we can perform this procedure in the kernel
(\ref{K^{Jneq0}_{2nu}(|x|<eta,eta|y)-dual1})
or its Mellin-Barnes integral
representation
(\ref{K_{2nu}(x,eta|y)-MBb}),
\begin{eqnarray}
K^{J\neq0}_{2\nu}(-x|y) &\!\!\! = \!\!\!& \frac{1}{[y^{2\nu}]_+}\int_{-1}^{1}\!\frac{dz}{\pi}\,
\frac{i^{1-2\nu} \left(1-x^2\right)}{2\left[(i y z +x)^2+1-x^2\right]^{\frac{3}{2}}} \,
\frac{z^{1-2 \nu }}{\sqrt{1-z^2}}+ {\rm c.c.}
\\
 &\!\!\!=\!\!\!&  \frac{1}{2i}\int^{c+i \infty}_{c-i \infty}\!dj\,
\frac{2^{2\nu-1}\,p_{j+2\nu}(-x)}{2\sin([j+1]\pi/2)} \frac{\Gamma(3+j+2\nu)\Gamma\!\left(\frac{2+j}{2}\right)}{\Gamma\!\left(\frac{5}{2}+j+2\nu\right)\Gamma\!\left(\frac{3+j}{2}\right)}\; y^{j}\,,
\nonumber
\end{eqnarray}
where $K^{J\neq0}_{2\nu}(x|y)=\lim_{\eta\to \infty}K^{J\neq0}_{2\nu}(x\eta,\eta|y)$.
Adding the $J=0$ term, we find
\begin{subequations}
\label{d(x,t)}
\begin{eqnarray}
\label{d(x,t)-dual}
d(x,t)  &\!\!\!=\!\!\!& \sum_{\nu=0}^\infty   \int_0^1\!dy \left[ K^{J\neq0}_{2\nu}(x|y)- K^{J\neq0}_{2\nu}(-x|y)  +
 2\left(\!1-x^2\right) C^{\frac{3}{2}}_{2\nu-1}(x) \frac{1}{y} \right]y^{2\nu} Q_{2\nu}(y,t)
\qquad\qquad
\\
{\ }\nonumber\\
\label{d(x,t)-MB}
&\!\!\!=\!\!\!& \sum_{\nu=0}^\infty \Bigg[
\frac{1}{2i}\int^{c+i \infty}_{c-i \infty}\!dj\,
\frac{{\rm sign}(-x)\,p_{j+2\nu}(-|x|)}{2^{j+1} \sin\!\left([j+1]\frac{\pi}{2}\right)}\, \frac{\Gamma(2+j)\Gamma\!\left(\frac{2+j}{2}\right)}{\Gamma\!\left(\frac{3}{2}+j\right)\Gamma\!\left(\frac{3+j}{2}\right)}\;
H^{}_{j+2\nu,j+1}(t)
\\
&&\phantom{\sum_{\nu=0}^\infty \Bigg[}
-2 p_{2\nu-1}(x) H_{2\nu-1,0}(t)
\Bigg].
\nonumber
\end{eqnarray}
\end{subequations}
Here we used antisymmetry to rewrite the Mellin-Barnes integral in such a manner that it converges fast
(at $j\to \infty$ for $|{\rm arg}(j)| \le \pi/2$).
Thus, we can conclude that the $D$-term extracted within the dual parametrization framework
coincides with that from the Mellin-Barnes integral approach.

\section{Elementary amplitude and ${\rm SO}(3)$-PW decomposition}
\label{sec:Amplitude}
\setcounter{equation}{0}

\subsection{Elementary amplitude }
The DVCS and DVMP amplitudes within the collinear factorization approach
are obtained by the convolution of GPDs with the hard-scattering parts given by the
appropriate partonic propagators. At the LO, the convolution formula
for the charge even Compton form factor (CFF) in the flavor singlet
sector involves the following elementary amplitude
\begin{equation}
\label{H^S}
{\cal
H}^{(+)}(\xi,t) = \int_{0}^{1}\!dx \left[\frac{1}{\xi-x-i
\epsilon} -\frac{1}{\xi+x-i\epsilon}\right]
H^{(+)}(x,\eta=\xi,t),
\end{equation}
where, in analogy to our GPD nomenclature,
${\cal H}^{(+)}$
stands for the charge even
($C=+1$) amplitude arising from the
unpolarized quark GPD combination
(\ref{Def_singlet_and_nonsinglet_combinations})
of a particular quark flavor.
Note that  the fractional quark charge squared factors are
not included in our elementary amplitude definition (\ref{H^S}).

\bi
\item
The imaginary part of the elementary amplitude
(\ref{H^S})
is given by the charge even GPD combination on the cross-over line
\be
{\rm Im} {\cal H}^{(+)}(\xi,t)=\pi H^{(+)}(\xi,\xi,t).
\ee

\item The real part of the amplitude
(\ref{H^S})
can be reconstructed from the known imaginary part with the
help of once subtracted (signature odd) dispersion relation,
\begin{eqnarray}
\label{DR-LO}
 {\rm Re} {\cal H}^{(+)}(\xi,t)=
{\cal P}\!\!\int_0^{1}\!dx\,\frac{2x\,
H^{(+)}(x,x,t)}{\xi^2-x^2
}  + 4 D(t)\,,
\end{eqnarray}
where
${\cal P}$
stands for the principle value regularization prescription. Note that the equivalence of this dispersion relation with
the convolution formula (\ref{H^S}) has been shown in \cite{Teryaev:2005uj,Anikin:2007yh} and that the dispersion relation has
been utilized to find the convolution formula from the operator product expansion in the unphysical region \cite{Chen:1997rc,Kumericki:2007sa}.

\item
The subtraction constant in the dispersion relation, the so-called
$D$-form factor,
results from the convolution of the $D$-term
(\ref{Def_D_term}),
\be
4 D(t)= \int_{0}^1\!dx\,\frac{2x\, d(x,t)}{1-x^2}\,,
\label{Def_D_form_factos}
\ee
which can be extracted from a particular GPD representation by means
of the `low-energy' limit
$\eta \to \infty$.
\ei

Our present goal is to compare the leading order expressions for the CFF in the dual
parametrization approach to that within the Mellin-Barnes-integral representation framework.
Instead of directly  using the convolution formula
(\ref{H^S})
we employ the dispersive approach making use of the dispersion relation
(\ref{DR-LO}).
Indeed, we already checked that the dual parametrization and the Mellin-Barnes-integral
framework provide the same expressions for the GPD on the cross-over line and, hence, the identical
absorptive parts of the amplitude.
Therefore, it only remains to check that both approaches produce the same
value of the $D$-form factor. The real part of the amplitude must then coincide in both representations, which we will  exemplify, too.
Moreover, it turns out that the expressions for the real part have a similar  mathematical structure.

Within the dual parametrization approach, employing the expression
(\ref{H+(x,x)-dual})
for the  GPD on the cross-over line, the dispersion relation
(\ref{DR-LO})
can be expressed  in terms of the convolution kernels as
\begin{eqnarray}
\int_0^{1}\!\frac{dx}{\pi}\,\frac{2x\,
K_{2\nu}(x,x|y)}{\xi^2-x^2-i\epsilon} = i K_{2\nu}(\xi,\xi|y) +
\frac{2\theta(1-\frac{2y}{\xi} + y^2)}{y\sqrt{1-\frac{2y}{\xi} + y^2}} +
\frac{2}{y\sqrt{1+\frac{2y}{\xi} + y^2}} - \frac{4}{y\sqrt{1+y^2}}\,.
\qquad
\label{Kernel_Real_Part_Dual_parametrization}
\end{eqnarray}
The $D$-form factor  can be computed by plugging the kernel
(\ref{d(x,t)-dual})
into the convolution formula
(\ref{Def_D_form_factos}).
This provides the familiar formal expression
\cite{Polyakov:2007rv,Polyakov:2008aa}
\begin{eqnarray}
4 D(t) 
&\!\!\!= \!\!\!&
4 \int_{0}^1\!\frac{dy}{y}\left[ \frac{1}{\sqrt{1+y^2}}-1\right] N(y,t)
+
4 \int_{0}^1\!\frac{dy}{y}   \left[ N(y,t)-Q_0(y,t) \right]\,,
\qquad
\label{D_ff_dual_parametrization}
\end{eqnarray}
where
$N(y,t)$
stands for the GPD quintessence function
(\ref{GPD_quintessence}).

Combining
(\ref{Kernel_Real_Part_Dual_parametrization}) and
(\ref{D_ff_dual_parametrization})
provides the well known result for the CFF in the LO
approximation within the dual parametrization approach
\begin{eqnarray}
\label{H^S-dual}
{\cal H}^{(+)}(\xi,t) &\!\!\!= \!\!\!& \int_{0}^1\!\frac{dy}{y}\!
\left[\frac{2}{\sqrt{1-\frac{2y}{\xi} + y^2 - i \epsilon}} +
\frac{2}{\sqrt{1+\frac{2y}{\xi} + y^2}}-4\right]
N(y,t)
\nonumber
\\
&&+ 4 \int_0^1\!\frac{dy}{y}
\left[ N(y,t)-Q_0(y,t) \right]\,.
\end{eqnarray}
Note that the second integral in the r.h.s.\ of
Eq.~(\ref{D_ff_dual_parametrization})
corresponds to the
$J=0$
cross channel exchange contribution into the amplitude and,
as we have discussed in
Sec.~\ref{sub_sec_cov_kernels},
requires a suitable regularization.

To work out the expression for the elementary amplitude
(\ref{H^S})
within the Mellin-Barnes integral we make use of
the dispersion relation for the Legendre functions%
\footnote{The last term on the r.h.s.~of
(\ref{d^{j+1}_{00}(x)-DR})
can be dropped, since after the integration over
$j$ in
(\ref{H^S-MB}) it produces zero.}
\begin{eqnarray}
\label{d^{j+1}_{00}(x)-DR}
\int_0^{1}\!\frac{dx}{\pi}\,\frac{2x\, x^{-j-1}
\hat{d}^{j+1}_{00}(x)}{\xi^2-x^2-i\epsilon} &=&
\xi^{-j-1} \left[i+\tan\!\!\left(\!\frac{\pi j}{2}\!\right)\right]
\hat{d}^{j+1}_{00}(\xi) +
\frac{2^{-j-1}}{\sin\!\left([j+1]\frac{\pi}{2}\right)}\,
\frac{\Gamma(2+j)\Gamma\!\left(\frac{2+j}{2}\right)}{\Gamma\!\left(\frac{3}{2}+j\right)\Gamma\!\left(\frac{3+j}{2}\right)}
\phantom{\Bigg)}
\nonumber\\
&&\qquad + \frac{2^{-2j-2}\Gamma(2+j)^2\,\xi^{2+j}
}{\Gamma\!\left(\frac{3}{2}+j\right)
\Gamma\!\left(\frac{5}{2}+j\right)}\,
{_2F_1}\!\!\left(\!{(2+j)/2,(3+j)/2\atop (5+2j)/2}\Big|\xi^2\!\right).
\phantom{\Bigg)}
\end{eqnarray}

The integral representation of the subtraction constant, obtained
from the $D$-term expression
(\ref{d(x,t)-MB}),
reads
\begin{eqnarray}
\label{D(t)-MB}
4 D(t) 
 &\!\!\!= \!\!\!&
\frac{1}{2i}\int^{c+i \infty}_{c-i
\infty}\!dj\frac{(-1)}{\sin\!\left([j+1]\frac{\pi}{2}\right)}\sum_{\nu=0}^\infty
 \frac{2^{2\nu}\Gamma\!\left(\frac{5}{2}+j+2\nu\right)}{\Gamma\left(\frac{3}{2}\right)\Gamma(3+j+2\nu)}\, \frac{\Gamma(2+j)\Gamma\!\left(\frac{2+j}{2}\right)}{\Gamma\!\left(\frac{3}{2}+j\right)\Gamma\!\left(\frac{3+j}{2}\right)}\;
H^{}_{j+2\nu,j+1}(t)
\nonumber\\
&& + 2\sum_{\nu=1}^\infty \frac{2^{2 \nu } \Gamma\!\left(\frac{3}{2}+2
\nu\!\right)}{\Gamma\!\left(\frac{3}{2}\!\right)\Gamma(2+2
\nu)}H_{2\nu-1,0}(t).
\end{eqnarray}
Using
(\ref{HjJ2B})
one might convert the Mellin-Barnes integral expression for the $D$-form factor
(\ref{D(t)-MB})
into the series
\begin{eqnarray}
4 D(t) 
 &\!\!\!= \!\!\!& 4
\sum_{J=2 \atop {\rm even}}^\infty
 \frac{(-1)^{\frac{J}{2}}
\Gamma\!\left(\frac{1+J}{2}\right)}{\Gamma\!\left(\frac{1}{2}\right)
\Gamma\!\left(\frac{2+J}{2}\right)} \;\sum_{\nu=0}^\infty
B_{J+2\nu-1,J}(t)
+ 4\sum_{\nu=1}^\infty  B_{2\nu-1,0}(t)\,.
\end{eqnarray}
It is straightforward to check that
(\ref{D(t)-MB})
coincides with the dual parametrization result
(\ref{D_ff_dual_parametrization}).

Employing the expression for the
GPD on the cross-over line
(\ref{H+(x,x)-MB})
and combining
(\ref{d^{j+1}_{00}(x)-DR})
and
(\ref{D(t)-MB}),
we establish the Mellin-Barnes integral representation for the  CFF
\begin{eqnarray}
\label{H^S-MB}
{\cal H}^{(+)}(\xi,t)&\!\!\!=\!\!\!& \frac{1}{2i}\int^{c+i
\infty}_{c-i \infty}\!dj\, \left(\!\frac{2}{\xi}\!\right)^{j+1}
\!\left[i+\tan\!\!\left(\!\frac{\pi j}{2}\!\right)\right]
\hat{d}^{j+1}_{00}(\xi)
\sum_{\nu=0}^\infty
\frac{2^{2\nu}\Gamma\!\left(\frac{5}{2}+j+2\nu\right)}{\Gamma\left(\frac{3}{2}\right)\Gamma(3+j+2\nu)}
 H^{}_{j+2\nu,j+1}(t)
\nonumber\\
&&+
4\sum_{\nu=1}^\infty \frac{2^{2\nu}
\Gamma\!\left(\frac{3}{2}+2\nu\right)}{\Gamma\!\left(\frac{1}{2}\right)\Gamma(2+2\nu)}
H^{}_{2\nu-1,0}(t)\,.
\end{eqnarray}

Obviously, the structure of the two expressions
(\ref{H^S-dual})
and
(\ref{H^S-MB})
for the LO elementary amplitude matches.
Therefore, given that within the two approaches the elementary amplitude
satisfies the same once subtracted  dispersion relation
(\ref{DR-LO})
with identical absorptive parts, and the values of the corresponding subtraction constants
coincide, we conclude that the  Mellin-Barnes integral  and the dual parametrization
approaches result in identical expressions for the elementary amplitude
(\ref{H^S}).

Note that this expression  for the elementary amplitude within the
Mellin-Barnes integral approach
(\ref{H^S-MB})
now explicitly includes the $J=0$
cross-channel  exchange contribution
\begin{eqnarray}
4\sum_{\nu=1}^\infty \frac{2^{2\nu}
\Gamma\!\left(\frac{3}{2}+2\nu\right)}{\Gamma\!\left(\frac{1}{2}\right)\Gamma(2+2\nu)}
H^{}_{2\nu-1,0}(t)=
4\sum_{\nu=1}^\infty  B_{2\nu-1,0}(t)=
4\int_{0}^1\!\frac{dy}{y}   \left[ N(y,t)-Q_0(y,t) \right]
\end{eqnarray}
omitted in the phenomenological
application of
Ref.~\cite{Kumericki:2009uq}, see footnote
\ref{note:subtraction_term}.

\subsection{Calculation of the $D$-form factor}
\label{ssec:D-ff}
In this subsection we address the problem of assigning meaning
to the so-far formally written integral representation for the
$D$-form factor
(\ref{D_ff_dual_parametrization})
in the framework of the dual parametrization and discuss
the possible $J=0$ fixed pole
contribution into this quantity.

Once we adopt our usual assumptions on the small-$y$ asymptotic behavior
of the forward-like functions
($y^{2 \nu} Q_{2 \nu}(y,t) \sim y^{-\alpha(t)}$,
where
$\alpha(t)$
is the leading Regge trajectory),
the inverse momentum integrals in
(\ref{D_ff_dual_parametrization}),
(\ref{H^S-dual})
require suitable regularization.
This returns us to the discussion of Sec.~\ref{sub_sec_cov_kernels}
on the analytic properties of dPWAs and possible
$J=0$
fixed pole contributions.
Following the reasoning of Sec.~\ref{sub_sec_cov_kernels},
we assume that the functions
$y^{2 \nu} Q_{2 \nu}(y,t)$
and the GPD quintessence function
$N(y,t)$
belong to a sufficiently good class of functions
(see discussion in Ref.~\cite{Polyakov:2008aa})
so that  the resulting dPWAs
$B_{nJ}(t)$
turn to be analytic functions of $J$ up to a possible $J=0$ fixed pole contribution
manifest as the Kronecker-delta singularity $\sim \delta_{J0}$.
This allows to assign meaning to the formal expression for the
$D$-form factor (\ref{D_ff_dual_parametrization}):
\be
4 D(t)
&\!\!\!= \!\!\!&
4 \int_{0}^1\!\frac{dy}{y}\left[ \frac{1}{\sqrt{1+y^2}}-1\right] N(y,t)
+
4 \int_{(0)}^1\!\frac{dy}{y}   \left[ N(y,t)-Q_0(y,t) \right] + 4 D^{\rm f.p.}(t)\,,
\qquad
\label{D_ff_dual_parametrization_regularized}
\ee
where $(0)$ stands for the analytic regularization prescription and
\begin{eqnarray}
\label{D^{f.p.}(t)}
D^{\rm f.p.}(t) = \sum_{\nu=1}^\infty \frac{2^{2 \nu }
\Gamma\!\left(\frac{1}{2}+2
\nu\!\right)}{\Gamma\!\left(\frac{3}{2}\!\right)\Gamma(2+2 \nu)}H^{\rm
f.p.}_{2\nu-1,0}(t) = \sum_{\nu=1}^\infty  B^{\rm f.p.}_{2\nu-1,0}(t)\,
\end{eqnarray}
corresponds to the possible
$J=0$
fixed pole contribution into the $D$-form factor.

As pointed out in
Ref.~\cite{Polyakov:2008aa},
the expression for the $D$-form factor
(\ref{D_ff_dual_parametrization_regularized})
is a particular realization of the so-called GPD sum rule, derived from dispersion and
operator product expansion techniques, that to the LO accuracy reads
\cite{Kumericki:2008di}
\footnote{A similar looking inverse momentum sum rule was also formally derived in
\cite{Anikin:2007yh} by employing the
polynomiality constraints of GPDs and taking the $\xi\to 0$  limit.
}:
\be
\label{Inverse_momentum_SR}
2 D(t)= \lim_{j \rightarrow -1} \int_{(0)}^1 dx \, x^j \left[
H^{(+)}(x,x,t)-H^{(+)}(x,0,t)
\right] + 2 D^{\rm f.p.}(t)\,.
\ee
In
(\ref{Inverse_momentum_SR})
it is assumed that the $j$-th Mellin moment of
$H^{(+)}(x,x,t)-H^{(+)}(x,0,t)$
is an analytic functions of $j$
up to a possible
$j=-1$
fixed pole contribution. This analyticity assumption
is in fact equivalent to our $J$-analyticity assumption
for dPWAs
$B_{J n}(t)$
(see discussion in
\cite{Khuri:1963zza}).

Thus, the sum rule
(\ref{Inverse_momentum_SR})
expresses
(\ref{DR-LO})
through the high energy asymptotic behavior of a GPD on the
cross-over line and  the   high energy asymptotic behavior of the corresponding $t$-dependent PDF
up to an eventual
$j=-1$
fixed pole contribution
$D^{\rm f.p.}(t)$.
Note that the possible existence of this latter contribution in
(\ref{Inverse_momentum_SR})
was originally pointed out only verbally.
Unfortunately, the presence of an arbitrary fixed pole contributions obviously
deprives the GPD sum rule
(\ref{Inverse_momentum_SR})
(or its particular realization in the dual parametrization framework
(\ref{D_ff_dual_parametrization_regularized}))
of any practical predictive power for fixing the
$D$-form factor.

It is interesting to note that the fixed pole contribution into the
sum rule
(\ref{Inverse_momentum_SR})
turns to be defined by the forward-like functions
$Q_{2 \nu}(y,t)$
with
$\nu\ge 1$.
Indeed, one can check that the
$j=-1$
fixed pole contribution in
(\ref{Inverse_momentum_SR})
is given as sum of the
$J=0$
fixed poles
(\ref{D^{f.p.}(t)}),
while the contribution of the inverse GPD  momentum
\be
&&
B_{-1,0}(t)={\rm Reg}\int_{0}^1 \frac{dy}{y}   Q_{0}(y,t)= \frac{1}{2} H_{-1,0}(t)
=  \frac{1}{2} {\rm Reg}\int_{0}^1  \frac{dx}{x} H^{(+)}(x,0,t)
\nonumber \\ &&
\equiv \frac{1}{2} \int_{(0)}^1 \frac{dx}{x}  H^{(+)}(x,0,t) + B^{\rm
f.p.}_{-1,0}(t) \nonumber
\ee
does not contribute into the sum rule
(\ref{Inverse_momentum_SR}).
Obviously, the inverse momentum of
$Q_{0}(y,t)$
is exactly canceled
in the second term  on the r.h.s.~of
Eq.~(\ref{D_ff_dual_parametrization_regularized}).
In this sense, the fixed pole contribution into the sum rules
(\ref{Inverse_momentum_SR}),
(\ref{D_ff_dual_parametrization_regularized})
turns to be the essentially non-forward effect defined
by $Q_{2 \nu}$ functions with $\nu>0$, which are not constrained in the
$\eta \to 0$ limit.

In its ultimate formulation (so-called ``analyticity of the second kind'' assumption
\cite{Alfaro_red_book})
the analyticity in
$J$
for DPWAs implies the absence of
$J=0$
fixed pole (or the same
$j=-1$
fixed pole in the sum rule
(\ref{Inverse_momentum_SR}))
contribution into the generalized form factors
$B_{2\nu-1 \, J}(t)$
(\ref{Pot_divergent_FFs}).
This assumption might be considered as
the additional  ``external principle'' that can be deliberately employed when
building GPD models.
Unfortunately, examples of consistent pion GPD models, for which  the
``analyticity of the second kind'' assumption
is not respected, are well known from effective theories
\cite{SemenovTianShansky:2008mp}.

Let us also point out that the use of analyticity for the evaluation of
the $D$-form factor brings some complications for model builders. For instance,
introducing the $t$-dependence of the forward-like functions  through
the leading Regge trajectory
$\alpha(t)$ implies that
the expression for the $D$-form factor turns to be divergent
in the specific cases when the Regge trajectory passes the integer
values zero and one
\cite{SemenovTianShansky:2010zv}.
This probably should be seen as an artifact of the common way for
introducing the
$t$
dependence of GPDs through the Regge trajectories that may turn to be
an oversimplification.
Therefore, for practical purpose of data description it seems to be  more
appropriate to abandon the proposal to fix the subtraction
constant in the dispersion relation from the high energy asymptotic
behavior thought the
inverse momentum GPD sum rule
(\ref{Inverse_momentum_SR})
implying the use of the analytic regularization. Instead,
{\it e.g.},
as done within the global GPD fitting procedure
\cite{Kumericki:2009uq},
the $D$-form factor can from the very beginning be treated as an independent subtraction
constant that has to be determined form the data analysis.

\subsection{SO(3)-PW decomposition, Froissart-–Gribov projection and $J=0$ fixed pole}
\label{sec: SO(3)PW and fixed pole}

Employing the once subtracted dispersion relation for the LO elementary
amplitude one can derive the extremely useful integral representation for the
corresponding
${\rm SO}(3)$
PWAs \cite{Kumericki:2008di}, known as the Froissart-–Gribov projection
\cite{Gribov1961,Froissart1961}.

For this issue we consider the dispersion relation for the elementary amplitude
${\cal H}^{(+)}$
analytically continued to the
$t$-channel:
\be
{\cal H}^{(+)}(\cos \theta_t,t)=\int_{0}^1\!dz\, \frac{2z}{1-z^2} \Phi^{(+)}(z,\cos\theta_t,t)  = \int_{0}^1\!
dx \frac{2x \cos^2\theta_t}{1-x^2 \cos^2\theta_t} H^{(+)}(x,x,t) + 4 D(t)\,,
\label{Elamp_crossed_disp}
\ee
where
$\Phi^{(+)}(z,\omega,t)=
H^{(+)}\!\left(\frac{z}{\omega},\eta=\frac{1}{\omega},t\right)$
stands for the charge even generalized distribution amplitude.
To deal with the cross channel
${\rm SO}(3)$-PWAs
\be
a_{J}
(t) \equiv \frac{1}{2}\int_{-1}^{1}\!d(\cos\theta_t)\,
P_J(\cos\theta_t)  {\cal H}^{(+)}(\cos \theta_t,t)
\label{a_J(t)_def}
\ee
we introduce the generalized distribution
amplitudes with a definite angular momentum $J$
\be
\Phi_J^{(+)}(z,t)=\frac{1}{2}\int_{-1}^{1}\!d(\cos\theta_t)\,
P_J(\cos\theta_t)\, \Phi^{(+)}(z,\cos\theta_t,t).
\ee
Employing the dispersion relation
(\ref{Elamp_crossed_disp})
together with Neumann's integral representation for the
Legendre functions of the second kind
$ {\cal Q}_J$
with integer
$J \ge 0$
\cite{Gradshteyn}
\be
\frac{1}{2} \int_{-1}^1 dz P_J(z) \frac{1}{ z'-z} = {\cal Q}_J(z')
\ee
we establish the Froissart-–Gribov projection formula for the
cross channel ${\rm SO}(3)$-PWAs $a_{J}
(t)$.
For even positive $J$ it reads \cite{Kumericki:2007sa}:
\begin{eqnarray}
&&
a_{J>0}
(t)=  \int_{0}^1 dz \frac{2z}{1-z^2} \Phi_J^{(+)}(z,t) = 2 \int_0^1\! dx
\frac{{\cal Q}_J(1/x)}{x^2} H^{(+)}(x,x,t)\,.
\label{Froissart-–Gribov}
\end{eqnarray}
For $J=0$ we obtain
\be
a_{J=0}
(t)= 2 \int_0^1\! dx \left[\frac{{\cal Q}_0(1/x)}{x^2} -
\frac{1}{x}\right] H^{(+)}(x,x,t) + 4 D(t)\,.
\label{Froissart-–Gribov_J=0}
\ee
Note that, under our usual assumptions on the small-$x$ asymptotic behavior
of the charge even GPD/GDA on the cross over line ($H^{(+)}(x,x,t) \sim x^{-\alpha(t)}$ with $\alpha(0)>-2$),
both the
(\ref{Froissart-–Gribov})
and
(\ref{Froissart-–Gribov_J=0})
provide rigorous and finite results since
$\frac{{\cal Q}_J(1/x)}{x^2} \sim x^{J-1}$ for small $x$
and the $\frac{1}{x}$-term in the square brackets in
Eq.~(\ref{Froissart-–Gribov_J=0}) cancels the
$1/x$-singularity in $\frac{{\cal Q}_0(1/x)}{x^2} $, leaving a term of order
${\cal O}(x)$.

The ${\rm SO}(3)$-PWAs
(\ref{a_J(t)_def})
have also  been evaluated in the dual
parametrization framework by mapping the imaginary part of the amplitude
to the GPD quintessence function and calculating its Mellin moments
\cite{Polyakov:2007rv}.
In fact, to the leading order accuracy the GPD quintessence function
(\ref{GPD_quintessence})
contains exactly the same information as the GPD on the cross over line
but casted in terms of the dual parametrization auxiliary variable $y$.
As pointed out in \cite{Polyakov:2007rv},  GPD quintessence can be recovered from
the GPD on the cross over line  by the inverse Abel tomography procedure
yielding
\be
N(y,t)=
\frac{y(1-y^2)}{(1+y^2)^{\frac{3}{2}}}
\int_{\frac{2y}{1+y^2}}^1
\frac{d x}{x^{\frac{3}{2}}}
\frac{1}{\sqrt{x- \frac{2y}{1+y^2}}}
\left\{
\frac{1}{2}
H^{(+)}(x,x,t)
-
x \frac{d}{d x}
H^{(+)}(x,x,t)
\right\}\,.
\label{N(y)main}
\ee

Now the $(J-1)$-th Mellin moments of the
GPD quintessence function can be computed
employing the inverse Abel transform formula
(\ref{N(y)main})
as
\be
\int_0^1\! dy\, y^{J-1}  N(y,t) = \int_0^1\!dx\,
\left[\frac{1}{\sqrt{x}} \frac{d}{dx} R_J(x)\right] H^{(+)}(x,x,t)\,,
\ee
where the auxiliary functions
$R_J(x)$
can be expressed%
\footnote{Note that the original expression for
$R_J(x)$  Eq.~(28) of Ref.~\cite{Polyakov:2007rv}
(or Eq.~(26) for the ArXive version of Ref.~\cite{Polyakov:2007rv})
misses a factor $\frac{1}{\sqrt{2}}$.}  through the Legendre functions of the second kind
\be
\frac{1}{\sqrt{x}} \frac{d}{dx}  R_J(x)= \frac{1}{\sqrt{x}}
\frac{d}{dx}
\int_0^x\frac{dw}{ \sqrt{2} \sqrt{w} }\left(\frac{1-\sqrt{1-w^2}}{w}\right)^{J+\frac{1}{2}}
\frac{1}{\sqrt{x-w}} = \frac{\frac{1}{2}+J}{2}
\,\frac{2{\cal Q}_J(1/x)}{x^2}\,.
\ee
Finally, this allows to work out the expressions for
${\rm SO}(3)$-PWAs
(\ref{a_J(t)_def})
in the dual parametrization framework.
For even $J>0$ we get
\be
a
_{J>0}(t)=
\frac{4}{2J+1}   \sum_{n=J-1 \atop {\rm odd}}^\infty B_{n J}(t)=
\frac{4}{2J+1} \int_0^1 dy y^{J-1} N(y,t).
\label{a_J(t)_Dual}
\ee
For
$J=0$
it reads
\be
a
_{J=0}(t) &\!\!\!=\!\!\!&  4\sum_{n=1 \atop {\rm odd}}^\infty B_{n 0}(t)=
4  \, {\rm Reg} \int_0^1 \frac{dy}{y}   \left( N(y,t)-Q_0(y,t) \right)
\nonumber \\
&\!\!\!=\!\!\!&
4   \int_{(0)}^1 \frac{dy}{y}   \left( N(y,t)-Q_0(y,t) \right) + 4 D^{\rm f.p.}(t),
\label{a_J=0(t)_Dual}
\ee
where the eventual
$J=0$
fixed pole contribution into the $D$-form factor is defined in
Eq.~(\ref{D^{f.p.}(t)}).
As pointed out in
\cite{Polyakov:2007rv},
these results enlighten the physical meaning of the GPD quintessence function
$N(y,t)$.
Its Mellin moments carry valuable information about the hadron structure encoding how the target hadron responses
to the well-defined quark-antiquark probe of the particular spin
$J$.
One can check that  the dual parametrization expressions for the
${\rm SO}(3)$-PWAs
(\ref{a_J(t)_Dual}),
(\ref{a_J=0(t)_Dual})
are fully equivalent to the general Froissart-–Gribov projection formulas
(\ref{Froissart-–Gribov}),
(\ref{Froissart-–Gribov_J=0}).

The  Froissart-–Gribov projection allows for a  clear formulation of the
$J=0$
fixed pole issue that goes along with the text book view
\cite{MarSpea70}.
Employing  the analytic regularization, we separate two finite integrals in
the rigorously defined expression for
$J=0$
PW
(\ref{Froissart-–Gribov_J=0}):
\begin{eqnarray}
a_{J=0}
(t)= 2 \lim_{J\to 0} \int_{(0)}^1\! dx\,\frac{{\cal
Q}_J(1/x)}{x^2} H^{(+)}(x,x,t)-2 \lim_{j\to -1} \int_{(0)}^1\! dx\, x^j
H^{(+)}(x,x,t)  + 4 D(t)\,.
\label{a_{J=0}(t)-reg}
\end{eqnarray}
We emphasize that this equality should be considered as valid in the
presence of
$j=-1$
or
$J=0$
fixed poles for DPWAs. Similarly to our analysis of the inverse momentum sum rule
(\ref{Inverse_momentum_SR}),
we consider the
$J=0$
fixed pole contribution to the $J=0$ PW $a_{J=0}
$ as an unknown addendum to the analytic continuation of
$a_{J}
(t)$ to $J=0$. Therefore, we write the $J=0$ PWA as
\begin{eqnarray}
a_{J=0}
(t)=  \lim_{J\to 0} 2 \int_{(0)}^1\! dx\, \frac{{\cal
Q}_J(1/x)}{x^2} H^{(+)}(x,x,t) +  a^{\rm f.p.}_{J=0}(t)\,.
\label{a_{J=0}(t)-dec}
\end{eqnarray}
Inserting the inverse momentum sum rule
(\ref{Inverse_momentum_SR})
and
(\ref{a_{J=0}(t)-dec})
into the equality
(\ref{a_{J=0}(t)-reg}),
we find that the
$J=0$
and
$j=-1$
fixed pole contributions differ
by the analytically regularized inverse moment of the $t$-dependent PDF
\be
a^{\rm f.p.}_{J=0}(t)=-2 \int_{(0)}^1 \frac{dx}{x} H^{(+)}(x,0,t)
+4 D^{\rm f.p.}(t).
\label{Analytic_J=0_PW}
\ee
In the first term in the r.h.s.~of
(\ref{Analytic_J=0_PW})
we immediately recognize the universal local two-photon-to-quark-coupling
$J=0$ pole contribution (corresponding to the ``seagull'' diagram'')
fiercely advocated in
\cite{Brodsky:2008qu,Szczepaniak:2007af}.
However, we would like to emphasize that this term provides the complete
$J=0$ fixed pole contribution into the DVCS amplitude only once
the ``analyticity of the second kind assumption'' requiring
$D^{\rm f.p.}(t)=0$ is valid. This was in fact assumed in Ref.~\cite{Brodsky:2008qu}.
Moreover, as explained in Sec.~\ref{ssec:D-ff}, the analytically regularized inverse $t$-dependent PDF momentum contribution
cancels in the DVCS amplitude.

It is worth reminding that the problem of the
$J=0$
fixed pole contribution was broadly discussed in the late sixties and in the seventies for the
forward Compton scattering
(see {\it e.g.}
Refs.~\cite{Gross:1968,Creutz:1968ds,Brodsky:1971zh,Brodsky:1973hm,Zee:1972nq,Creutz:1973zf}).
However, it was finally recognized
\cite{Creutz:1973zf}
that unfortunately ``{\it there exists no general theoretical
argument, independent of specific models, for such a singularity}'',
making it an optional model-dependent contribution.
We also can not point out the general theoretical justification
for the analyticity principle allowing to fix
$D^{\rm f.p.}(t)$.
As we explained in Sec.~\ref{ssec:D-ff},
the use of  the ``analyticity of the second kind assumption'' for GPD modeling is not mandatory.

\section{GPD model examples}
\label{sec:example}
\setcounter{equation}{0}

The important aspect in GPD fits to experimental data is the control over the normalization of the resulting amplitudes.
In particular, the normalization of the imaginary part plays the crucial role.
As explained in
Sec.~\ref{sec-cases_cross-over},
the normalization of the imaginary part in the leading order approximation, {\it i.e.},
the GPD on the cross-over line, is determined by the
sum over all ${\rm SO}(3)$-PWs (see the expression (\ref{H+(x,x)-dual})
through the GPD quintessence function (\ref{GPD_quintessence})
or the Mellin-Barnes integrand of Eq.~(\ref{H+(x,x)-MB}).
In the smaller-$\xB$ region, the amplitude turns to be proportional to the sum of residues
$\mathbb{R}^\alpha_\nu$ of the leading Regge trajectory $\alpha(t)$,
$$
{\cal H}(\xB,t,Q^2) \propto \sum_{\nu =0}^\infty  \mathbb{R}^\alpha_\nu(t,Q^2)\,.
$$
Therefore, to control the normalization it actually suffices to employ just two non-vanishing
$\nu=0$
and
$\nu=1$
contributions, where the
$\nu=1$
one might be considered as an `effective' one.
Clearly, without further model assumptions and neglecting perturbatively predicted
$Q^2$-evolution one can build an infinite number of GPDs that allow to describe experimental data for small-$\xB$.
However, in the small-$\xB$ region HERA collider data have a large
$Q^2$-lever arm,
$1 \, \GeV^2 \lesssim Q^2 \lesssim 100 \, \GeV^2 $.
Therefore, the
$Q^2$-evolution should be beyond doubts implemented in the data analysis.
Since for increasing
$\nu$
the partial  residues
$\mathbb{R}_\nu(t,Q^2)$
to the GPD are getting  more and more suppressed with growing
$Q^2$,
one can use three contributions with
$\nu\in\{0,1,2\}$
to control the normalization of the amplitude and its change with
$Q^2$, for a more detailed discussion see Refs.~\cite{Muller:2009zzd,Mueller:2011xd}.

In this section we have a closer look to a generic GPD model, set up in the  dual parametrization approach,
that includes $\nu \le 2$ contributions. We also present the Kumeri{\v c}ki-M{\"u}ller model, which arose from the 
description of experimental DVCS data, and convert it into  the dual parametrization framework.

\subsection{A generic model for the dual parametrization}
To get some insight into the dual parametrization, we adopt the following generic parametrization for all
forward-like functions at
$t=0$%
\footnote{This is a bit restrictive class of functions, but considering it as a building block one can easily
generalize the discussion {\it e.g.} to the class of functions
$y^{-\alpha} (1-y)^\beta(1+ \sum_i c_i y^{\gamma_i})$
familiar from PDF fits.}
,
\begin{eqnarray}
y^{2\nu} Q_{2\nu}(y,t=0) = n_{2{\nu}} \frac{5 \Gamma(3-\alpha +\beta)}{6 \Gamma(2-\alpha) \Gamma(1+\beta)}\, y^{-\alpha} (1-y)^\beta\,,
\label{ansatz-generic}
\end{eqnarray}
where only the normalization depends on
$\nu$.
It is chosen in such a manner that
$n_{0}\equiv M_2$
is the momentum fraction average of the corresponding PDF.
The (conformal) GPD moments of this model are easily calculated from
(\ref{HjJ2B}),
\begin{eqnarray}
\label{HjJ2B-toy}
 H^{(+)}_{jJ} = \frac{2^{J}\Gamma(3+j)\Gamma\!\left(\frac{1}{2}+J\right)}{2^{j+1} \Gamma\!\left(\frac{5}{2}+j\right)\Gamma(1+J)}
 B_{jJ}\,,\;
 B_{jJ}= n_{j+1-J}\frac{5 \Gamma(3-\alpha +\beta)\Gamma(J-\alpha)}{6 \Gamma(2-\alpha)\Gamma(1+J-\alpha +\beta)}\,.
\end{eqnarray}
Apart from the normalization factor
$n_{j+1-J}$,
the dPWAs
$B_{jJ}$
depend only on
$J$,
which implies that the
$j$- and $J$-dependencies in
$ H^{(+)}_{jJ}$
factorize.

Below we consider the important limiting cases for a
GPD within the dual parametrization toy model
(\ref{ansatz-generic}).

\begin{itemize}
\item{{\bf $\eta \to 0$ limit}}

The corresponding PDF  calculated from
(\ref{Q_02H(x,0,t)})
\begin{eqnarray}
\label{H^{(+)}-nfPDF}
H^{(+)}(x,0,t=0)&\!\!\!=\!\!\!& \frac{M_2 5 \Gamma(3-\alpha +\beta)}{6 \Gamma(2-\alpha) \Gamma(1+\beta)}
\Bigg\{x^{-\alpha }(1-x)^{\beta} + \frac{\Gamma\!\left(-\frac{1}{2}-\alpha\right) \Gamma(1+\beta)}{2\Gamma\!\left(\frac{1}{2}-\alpha +\beta\right)}\, x^{1/2}
\qquad\qquad\\
&&\phantom{N_{0}\frac{5 \Gamma(3-\alpha +\beta)}{6 \Gamma(2-\alpha) \Gamma(1+\beta)}}+ \frac{x^{-\alpha }}{1+2\alpha } \,
{_2F_1}\!\!\left(\!{-1/2-\alpha, -\beta \atop 1/2-\alpha}\bigg|x\!\right)\Bigg\}\,,
\nonumber
\end{eqnarray}
has a rather intricate functional form. It possesses the same small-$x$
asymptotic behavior as the input forward-like function, however, has a different normalization, which reads for
$\alpha > -1/2$
as following:
$$
\lim_{x\to 0} H^{(+)}(x,0,t=0)= \frac{2 (1+\alpha )}{1+2 \alpha } \lim_{x\to 0} Q_0(x,t=0)\,,\quad
\lim_{x\to 0} Q_0(x,t) =  \frac{5 M_2 \Gamma(3-\alpha +\beta)}{6 \Gamma(2-\alpha) \Gamma(1+\beta)}\,  x^{-\alpha }\,.
$$
Employing the behavior of
$_2F_1$-function in the vicinity of its branch point
$x=1$,
we find that the two functions coincide in the limit
$x\to 1$,
$$
\lim_{x\to 1} H^{(+)}(x,0,t=0)=  \lim_{x\to 1} Q_0(x,t=0)\,,\quad  \lim_{x\to 1} Q_0(x,t=0) =
M_2
\frac{5 \Gamma(3-\alpha +\beta)}{6 \Gamma(2-\alpha) \Gamma(1+\beta)}\, (1-x)^\beta\,.
$$

\item {\bf GPD on the cross-over line}

While for
$\eta=0$
the contribution of forward-like functions with
$\nu \ge 1$
vanishes, all forward-like functions contribute on the
same footing into GPD on the cross-over line
$x=\eta$.
Employing the integral  transform
(\ref{K_{2nu}(x,x|y)})
with the GPD quintessence function
(\ref{GPD_quintessence})
of the model
(\ref{ansatz-generic})
we conclude that the overall normalization of the corresponding  GPD on the cross-over line  is determined by the sum of the normalization factors
\be
&&
H^{(+)}(x,x,t=0)
\nonumber \\ &&
= \left[\sum_{\nu=0}^\infty n_{\nu}\right]
\frac{5 \Gamma(3-\alpha +\beta)}{6 \Gamma(2-\alpha) \Gamma(1+\beta)}\,\frac{1}{\pi}\int_{\frac{x}{1+\sqrt{1-x^2}}}^1\!\frac{dy}{y}\,
\sqrt{\frac{x}{y}}\frac{\sqrt{2}}{\sqrt{1-\frac{1+y^2}{2 y}x}}\,   y^{-\alpha} (1-y)^\beta\,.
\nonumber \\ &&
\label{Hxx_model}
\ee
Comparing the small-$x$ asymptotics of
(\ref{Hxx_model})
to that of the PDF,  for  each partial contribution we recover the familiar enhancement factor
(skewness ratio)
\cite{Shuvaev:1999ce},
given by the Clebsch-Gordon coefficient
$r^q=\frac{2^{\alpha}\Gamma\!\left(\frac{3}{2}+\alpha\right)}{\Gamma\!\left(\frac{3}{2}\right)\Gamma(2+\alpha)}$
of the conformal partial wave expansion
at
$j=\alpha-1$.
For the `reggeon'
($\alpha=1/2$)
and the `pomeron'
($\alpha=1$)
Ans\"{a}tzen the enhancement factor reads
\begin{equation}
r^q\big|_{\alpha=1/2}=\frac{8 \sqrt{2}}{3\pi}  \approx 1.2\,;
\ \ \ \ \ \
r^q\big|_{\alpha=1}= \frac{3}{2}.
\label{skewness-x_small}
\end{equation}
Consequently, the complete skewness ratio in the toy model
(\ref{ansatz-generic})
is adjustable and is given by
\begin{equation}
\lim_{x\to 0}\frac{H^{(+)}(x,\eta=x,t=0)}{H^{(+)}(x,0,t=0)} =
\left[1+\frac{\sum_{\nu=1}^\infty n_{\nu}}{n_0}\right]\frac{2^{\alpha}\Gamma\!\left(\frac{3}{2}+\alpha\right)}{\Gamma\!\left(\frac{3}{2}\right)\Gamma(2+\alpha)}\,.
\label{skewness-x_small_sum}
\end{equation}
The large $x$-behavior of the GPD on the cross-over line is  inherited from both the $\beta$-parameter and the chosen set of
${\rm SO}(3)$-PWs.
For the Legendre polynomials we have a
$(1-x)^\frac{\beta}{2}$
behavior, {\it i.e.},
the skewness ratio%
\footnote{Clearly, employing a wider class of functions $Q_{2\nu}$ the small-$x$ and large-$x$
skewness ratios can be made independent.}
\begin{equation}
\lim_{x\to 1} \frac{H^{(+)}(x,\eta=x,t=0)}{H^{(+)}(x,0,t=0)} \sim \left[1+\frac{\sum_{\nu=1}^\infty n_{\nu}}{n_0}\right] (1-x)^\frac{-\beta}{2}
\label{skewness-x_large}
\end{equation}
diverges in the
$x\to 1$
limit.

\item {\bf $D$-term}

Within the dual parametrization the $D$-term can be extracted  by means of the projection
(\ref{d(x,t)}). The $J>0$ contributions into  the $D$-term can be evaluated straightforwardly, while the
$J=0$ part
\begin{subequations}
\label{d^{J=0}-toy}
\begin{eqnarray}
d^{J=0}(x,t=0)  = \sum_{\nu=1}^\infty
 2\left(\!1-x^2\right) C^{\frac{3}{2}}_{2\nu-1}(x) B_{2\nu-1,0} 
\end{eqnarray}
involves the inverse moments of the forward-like functions (\ref{Pot_divergent_FFs})
that require suitable regularization.
Assuming the absence of the $J=0$ fixed pole contributions,
we employ the analytic regularization for the corresponding integrals:
\begin{eqnarray}
\label{B-toy-J=0}
 B_{2\nu-1,0} 
 &\!\!\!=\!\!\!& n_{2{\nu}} \frac{5 \Gamma(3-\alpha +\beta)}{6 \Gamma(2-\alpha) \Gamma(1+\beta)}\,  \int_{(0)}^1\!dy\, y^{-\alpha-1} (1-y)^\beta
\\
 &\!\!\!=\!\!\!& n_{2\nu}\frac{5 \Gamma(3-\alpha +\beta)\Gamma(-\alpha)}{6 \Gamma(2-\alpha)\Gamma(1-\alpha +\beta)}\,.
\nonumber
\end{eqnarray}
\end{subequations}
Note that in this case dPWAs
$B_{2\nu-1,0}$ possess poles for non-negative integer values of
$\alpha$. This implies that, once the Regge intercept is replaced by the Regge
trajectory $\alpha(t)= \alpha + \alpha^\prime t$,
it is unavoidable that poles appear in $B_{2\nu-1,0}(t)$ also for negative $t$ values
(see discussion in Sec.~\ref{sec: SO(3)PW and fixed pole}).
It might be cured, as suggested in Ref.~\cite{SemenovTianShansky:2010zv},
by modifying the residual $t$-dependence,
or, alternatively, by adding proper $J=0$ fixed pole contributions.
In the
$x\in [0,1]$
region the
$J=0$
part of the $D$-term possesses
$\nu$ nodes, which are inherited from the Gegenbauer polynomials
$C_{2\nu-1}^{3/2}$.
\end{itemize}

In the upper panels of
Fig.~\ref{fig:1}
we show the forward-like function
$x Q_0(x,t=0)$
of the toy model
(\ref{ansatz-generic})
as dashed curves with a `reggeon' (right panel) and `pomeron' (left panel) Ansatz, where we took generically the parameters
$$
n_0=M_2^{\rm val}=0.3,\; \alpha=1/2, \; \beta=3 \quad\mbox{and}\quad
n_0=M_2^{\rm sea}=0.15, \; \alpha=1, \; \beta=7\,,
$$
respectively.
The $\nu=0$ part of the GPD on the cross-over line
$H_0^{(+)}(x,x,t=0)$
is shown as solid curves.
It is enhanced with respect to PDF
$H^{(+)}(x,0,t=0)$
by the skewness ratios at small-$x$ and large-$x$, see
(\ref{skewness-x_small})
and
(\ref{skewness-x_large}),
respectively.
We emphasize again that the complete GPD,
which includes several forward-like functions
(\ref{ansatz-generic})
with  their normalization controlled by
$n_{2\nu}$,
is fully adjustable at small-$x$.
By taking higher forward-like functions with different values of the $\beta$-parameter,
the resulting GPD can be also set up flexible at
large-$x$.

The charge even GPD
$H^{(+)}(x,\eta,t)$,
can be numerically evaluated from the dual parametrization
(\ref{Q22H})
with the integral convolution kernels given in
(\ref{Kernel_with_J=0_explicit})--(\ref{Kernel_K_outer_region}),
and the forward-like functions
(\ref{ansatz-generic})
or, alternatively, from the Mellin-Barnes integral
(\ref{H^+(x,eta,t)-SO3-MB}) and the dPWAs (\ref{HjJ2B-toy}).
In the following we omit the $J=0$ contribution,
which requires to specify the fixed pole contributions
into the inverse moments (\ref{Pot_correct_FFs}).

In the lower left panel of
Fig.~\ref{fig:1}
we present the
$x$-shape of the `pomeron'-like GPD  at
$\eta=0.05$
with the three choices
$\nu\in \{0\ \mbox{(solid)} ,1\ \mbox{(dash-dotted)},2\ \mbox{(dashed)}\}$.
For clearness we took the same normalization for all conformal PWs
$n_0=n_{2}=n_{4}=M_2^{\rm sea}$.
One may notice, that the number of nodes
of $H_{2\nu}^{(+)}(x,\eta,t=0)$
in the central region is given by
$\nu$ and the suppression in the outer region increases with growing $\nu$.
Note that the GPD maximum is not located at the cross-over point $\eta=x=0.05$
but is rather  slightly shifted to the left. In the lower right panel
we display the corresponding  $D$-term in the region
$x\in[0,1]$ without the $J=0$
contribution (\ref{d^{J=0}-toy}).
Also here we observe that as in the central region and for the
$J=0$
contribution, see
(\ref{d^{J=0}-toy}),
$\nu$
nodes appear. The `reggeon'-like GPDs possesses the same qualitative features (not shown).

\begin{figure}[t]
\centering
\includegraphics[width=16 cm]{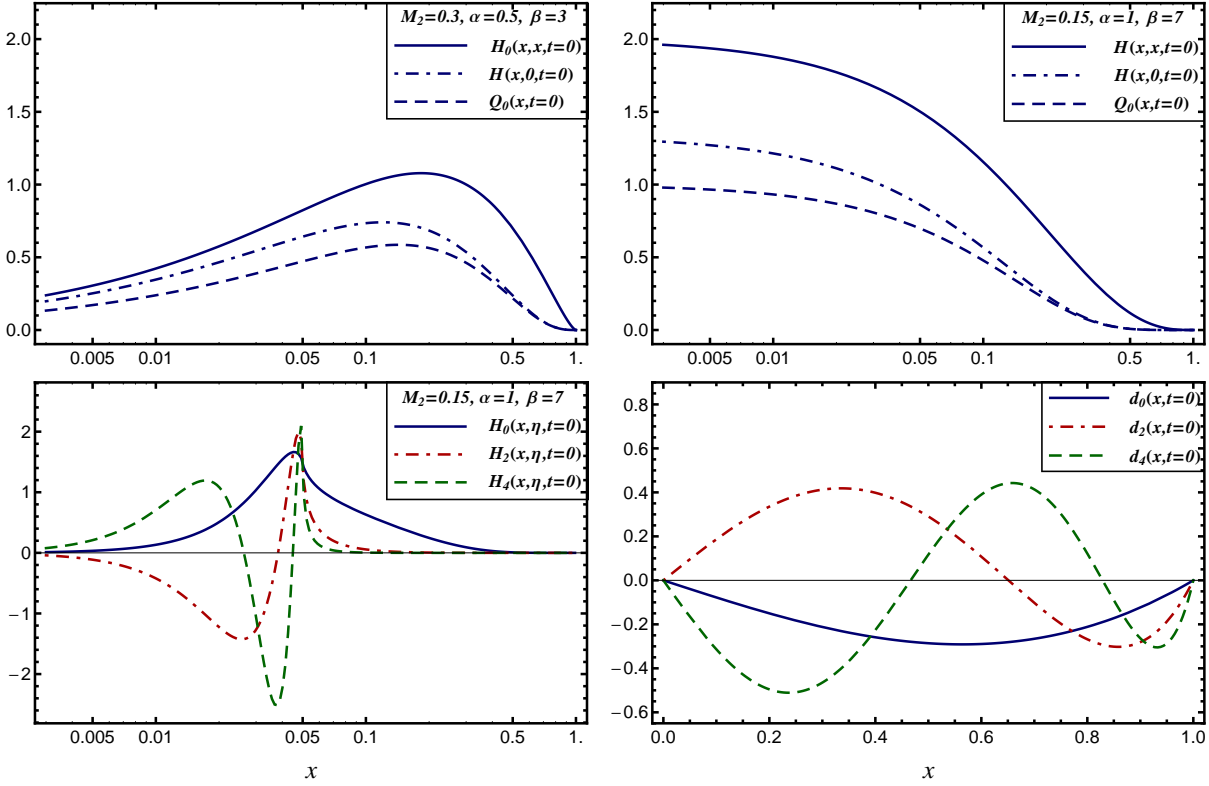}
\vspace{-4mm}
\caption{\small The upper panels show the generic model (\ref{ansatz-generic})
set up  in the dual parametrization:  forward-like functions
$Q_0(x,t=0)$ (dashed),  PDFs $H^{(+)}(x,0,t=0)$ (dash-doted),
and GPDs  on the cross-over line $H^{(+)}(x,x,t=0)$ (solid) for
$n_2=-n_1$ for a  `reggeon'-like ($M_2=0.3,\alpha=1/2,\beta=3$)
and `pomeron'-like ($M_2=0.15,\alpha=1,\beta=7$)
GPD model in the left and right panel, respectively. All functions are scaled with $x$
and are plotted versus the momentum fraction-type variable $x$.
In the lower  panel on the l.h.s.\ we show
the momentum fraction shape of the `pomeron'-like GPD, scaled with
$x$ at $\eta=0.05$. We take the same normalization for all partial waves ($n_0=n_2=n_4=0.15$)
and show the result for $\nu=0$ (solid), $\nu=1$ (dash-dotted), and
$\nu=2$ (dashed). On the r.h.s.~we show the corresponding $D$-term
as the function of $x$ for $\nu\in\{0,1,2\}$ within the same normalization.}
\label{fig:1}
\end{figure}

These more generic features of the dual parametrization can be qualitatively understood from the
properties of the dual parametrization convolution kernels, or directly from the
Mellin-Barnes integral. Let us first note that we can present the
$\nu>0$
pieces of the GPD by applying
$2\nu$
total derivatives in
$x$
on some auxiliary
function, in which the conformal PWs have index
$3/2+2\nu$,
$$
\eta^{2\nu}\, p_{j+2\nu}(x,\eta)  \propto \frac{d^{2\nu}}{dx^{2\nu}}
\oint_{(-1-\epsilon)}^{(+1+\epsilon)}\!du\,\frac{(u^2-1)^{j+1+2\nu}}{(x+\eta u)^{j+1}}\,.
$$
The increase of the suppression in the outer region with growing
$\nu$
can be traced back to the increasing number of total derivatives
that act on a monotonously decreasing function. This can be also seen on  the factor
$(\eta/x)^{2\nu}$
that appears in the conformal partial waves
(\ref{p_j(x,eta)-out}),
$$
\frac{\eta^{2\nu}p_{j+2\nu}(x\gg\eta, \eta)}{\sin(\pi[j+1])} \approx \frac{x^{-j-1}}{\pi}\, \left(\!\frac{\eta}{x}\!\right)^{2\nu},
$$
of the Mellin-Barnes integral
(\ref{H^+(x,eta,t)-SO3-MB}).
In the central region the GPD piece with
$\nu=0$
is a concave function, and consequently acting with total derivatives on it will generate nodes.
In the same way one can understand the functional form of the $D$-term contributions, shown in the lower right panel of
Fig.~\ref{fig:1},
which also possesses
$\nu$
nodes.
Also note that, since the
$\nu>0$
parts of the GPD contain
$2\nu$
total derivatives, their first
$2\nu$
Mellin moments vanish (we have numerically verified the polynomiality property  of our toy GPD by
evaluating the first few Mellin moments), while the non-vanishing ones
(with odd $N$) are given by
$$
\int_0^1\!dx\, x^{2\nu+1+N} H^{(+)}_{2\nu}(x,\eta,t=0) = \eta^{2 \nu +1 +N} \sum_{n=2\nu-1 \atop {\rm odd}}^{2 \nu +1 +N} B_{n, \, n-2\nu+1} (0)\,
P_{ n-2\nu+1}\!\left(\! \frac{1}{\eta}\! \right).
$$

\subsection{{\em KM10} model}

As explained above, in order to describe the present day experimental DVCS data
it suffices to consider the contribution of three first
${\rm SO}(3)$-PWs
(or, equivalently, within the dual parametrization framework, the contribution of three forward-like functions with
$\nu=0,\,1,\,2$).
For instance, in global DVCS fits the following
model, build up of three
${\rm SO}(3)$-PWs,
was employed for sea quarks at a input scale
$\mu^2=4\,\GeV^2$:
\begin{eqnarray}
\label{KM10-ansatz}
H^{\rm sea}_{j+2\nu,j+1}(t) &\!\!\!= \!\!\!&  M_2 s_{2\nu} \frac{B(1-\alpha+j,\beta+1)}{B(2-\alpha,\beta+1)}\,
\frac{1-\alpha+j}{1-\alpha(t)+j}\, \mbox{\boldmath $ \beta$}(t)\quad\mbox{for}\quad\nu= 0,\,1,\,2\,;
\\
H^{\rm sea}_{j+2\nu,j+1}(t) &\!\!\!= \!\!\!&  0\quad \mbox{for}\quad \nu > 2
\nonumber
\end{eqnarray}
with
$s_0=1$
and
$\alpha(t)= \alpha + \alpha^\prime t$.
This Ansatz is a part of the hybrid
Kumeri{\v c}ki-M{\"u}ller (KM)
model
\cite{Kumericki:2009uq}
that was pinned down by the global fit to the DVCS world data set for
unpolarized proton target
\cite{Kumericki:2010fr}
(for a detailed discussion see
Refs.~\cite{Kumericki:2011zc,Aschenauer:2013qpa}).
Thereby, the flexible Ansatz for sea quarks (and a similar one for gluons) enables one to describe the collider kinematics
data from the H1 and ZEUS collaborations to the leading order (or up to the next-to-next-leading) accuracy of perturbation theory,
with the residual $t$-dependence expressed by the dipole Ansatz
$\mbox{\boldmath $ \beta$}(t)= 1/(1-t/M^2)^2$.

The resulting {\em KM10} parameter set,
\begin{eqnarray}
\label{KM10-ps}
M_2=0.152,\;\; \alpha(t)=1.158+0.15 t,\;\; \beta=8,\;\; M^2=0.513\,\GeV^2, \;\; s_2=0.278,\;\; s_4=-0.130,
\qquad
\end{eqnarray}
is obtained from a
$\chi^2/{\rm d.o.f.} \approx 1$
fit to the deep inelastic scattering (DIS) and DVCS data. Here the normalization $M_2$ and the effective ``pomeron'' intercept are fixed
from the DIS data, the ``pomeron'' slope parameter $\alpha'$, the $\beta$ value and the cut-off mass,  as well as
the  {\em positive} $s_2$ and the {\em negative} $s_4$ values arise from the DVCS data.
Note that the fitting procedure fixes only the small-$\eta$
part of the model.

For simplicity, we do not employ the decomposition into the
electric and magnetic combinations of GPDs
(\ref{Electric_comb_GPDs}), (\ref{Magnetic_comb_GPDs})
and just perform the expansion in the ordinary Legendre polynomials
and make use of the Mellin-Barnes integral representation
(\ref{H^+(x,eta,t)-SO3-MB}),
where the
$J=0$
pole contributions are neglected.

The latter step is entirely legitimate and it corresponds to taking into account
the suitably chosen $J=0$ fixed poles contributions that exactly cancel the dPWAs
(\ref{KM10-ansatz})
at
$J=0$,
$$
H^{\rm sea}_{2\nu-1,0}(t) = M_2 s_{2\nu} \frac{(1-\alpha +\beta )(2-\alpha +\beta ) }{(\alpha -1)\alpha(t)}\,
\mbox{\boldmath $ \beta$}(t)\quad\mbox{for}\quad \nu= 1,\,2\,.
$$
This procedure also removes the unphysical pole at
$\alpha(t)=0$
that appears in
$H^{\rm sea}_{2\nu-1,0}(t)$
at rather large
$-t = \alpha/\alpha^\prime \sim 8\,\GeV^2$.
Note that in the {\em KM10} hybrid model fits the $D$-form factor is taken into account as an extra contribution.

Under the $Q^2$-evolution the ``pomeron'' intercept $\alpha$ will {\em effectively} increase for the
$\nu =0$
PW as it does for the $t$-dependent PDF.  Both
$\nu=1$
and
$ \nu=2$
PWs will evolve weaker than the
$\nu=0$
one. However, since of their alternating sign this evolution effect will partially cancel each other.

\begin{figure}[th]
\centering
\includegraphics[width=16. cm]{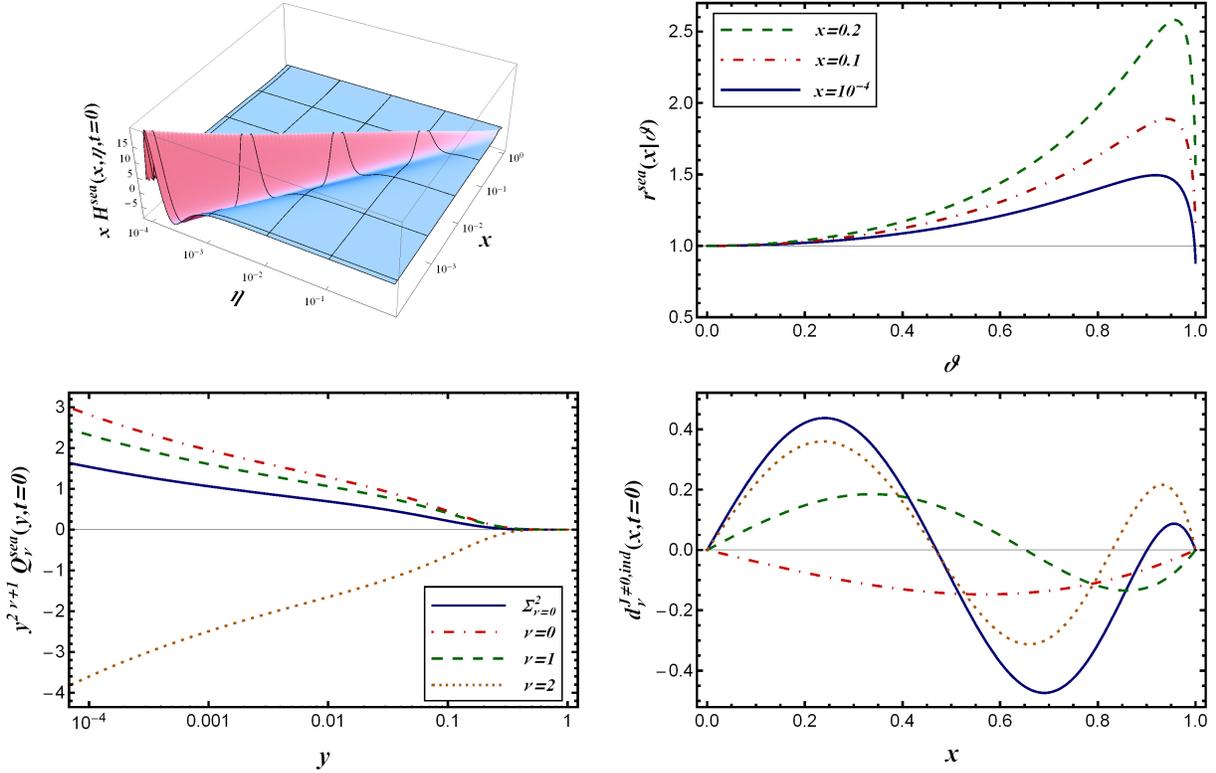}
\vspace{-4mm}
\caption{\small
The {\em KM10} GPD model, specified with the Ansatz
(\ref{KM10-ansatz})
and the parameter set
(\ref{KM10-ps}) for
$t=0$.
In the upper left panel the GPD as function of momentum fraction $x$ and skewness
$\eta$
is shown without
$J=0$
contribution, while in the upper right panel the skewness ratio as function of
$\vartheta$
for
$x=10^{-4}$
(solid),
$x=0.1$
(dash-doted) and
$x=0.2$
(dashed) is displayed. In the lower right panel the forward-like functions of the dual parametrization
are shown while on the lower left panel the induced $D$-term without
$J=0$
fixed pole is displayed, the solid line is the net contribution consisting of the
$\nu\in \{0\, \mbox{(dash-dotted)}, 1\, \mbox{(dashed)}, 2\, \mbox{(dotted)}\}$
contributions.
}
\label{fig-gpd3D}
\end{figure}
The resulting GPD
$H^{\rm sea}(x,\eta,t=0)$,
multiplied by
$x$,
is depicted at the input scale
$Q^2=4\,\GeV^2$
in the left upper panel of
Fig.~\ref{fig-gpd3D}
as a function of $x$ and $\eta$  for $t=0$.
The resulting GPD has nodes in the central region and also possesses a rather complex shape in the outer region.
To illustrate this in more detail, we plot the skewness ratio
$r(x|\vartheta)$
as a function of
$\vartheta=\eta/x$
for a small
$x$
value, taken to be
$10^{-4}$
(solid),
$x=0.1$
(dash-dotted) and
$x=0.2$
(dashed).
The nontrivial GPD shape in the outer region, for which
$0\le \vartheta\le 1$,
is now reflected by the concave shape of
$r(x|\vartheta)$
as function of
$\vartheta$.
In the small $x$-region the skewness ratio at
$\vartheta=1$
is a bit smaller than one and the concave shape  ensures that  it changes only little with growing
$Q^2$
in the experimentally accessible region. This stability is caused by a cancelation among the positive
second and negative third PW contribution while the first one evolves in the same manner as the
$t$-dependent PDF.
For
$\vartheta \gg 0$
the skewness ratio rapidly increases with growing
$x \gtrsim 0.2$.
According to the discussion in the preceding section this is caused by the more flat
large-$x$
behavior of the GPDs compared to that of the PDF. Note that at larger values of
$x$
the valence quark part starts to dominate.

We add that the {\em KM10} GPD model is qualitatively rather different from a GPD model build from a `standard' RDDA.
In this latter type of models the GPD in the outer region for fixed
$x$
turns to be a monotonically growing function of
$\eta$,
{\it i.e.},
$r(x|\vartheta)$
monotonously increases for
$\vartheta\in [0,1]$.
As a consequence of perturbative evolution, the skewness ratio at
$\vartheta=1$ will typically increase with growing $Q^2$, for examples see Ref.~\cite{Diehl:2007zu}.
It is known since more than one decade that to leading order accuracy in the collinear factorization approach
such a GPD model can not simultaneously describe the DVCS and DIS collider data from the H1 and ZEUS collaborations
\cite{Freund:2002qf}.

According to our explanations in the preceding section,  the non-trivial
$(x,\eta)$-shape
can be also easily understood in terms of the dual parametrization.
The inverse Mellin transform
(\ref{HjJ2Q})
allows us to map models build from conformal GPD moments into the space of forward-like functions.
For the model
(\ref{KM10-ansatz})
the inverse Mellin transform straightforwardly performed by employing the Cauchy theorem, for the partial contributions with
$\nu=0,\,1,\,2$
yields the following three forward-like functions:
\begin{eqnarray}
\label{KMreexpressed}
y^{2\nu} Q_{2\nu}(y,t)&\!\!\!=\!\!\! & M_2
\frac{s_{2\nu}\,\Gamma(3-\alpha +\beta)\,y^{-\alpha(t)}}{2^{-2 \nu+1}\Gamma(2-\alpha)\Gamma(\beta)}\, \mbox{\boldmath $ \beta$}(t)  \Bigg[
\frac{2\Gamma(1+\alpha(t))\Gamma\!\left(\frac{3}{2}+2 \nu +\alpha(t)\right)\Gamma(\beta)  }{
 \Gamma\!\left(\frac{1}{2}+\alpha(t)\right)  \Gamma(2+2 \nu +\alpha(t))\Gamma(1+\alpha^\prime t +\beta)}
\qquad \qquad \nonumber\\
&&
-
\frac{2\Gamma(\alpha)\Gamma\!\left(\frac{1}{2}+\alpha +2 \nu \right) y^{1+\alpha^\prime t}
}{
\Gamma\!\left(\alpha-\frac{1}{2}\right)\Gamma(1+\alpha +2 \nu) (1+\alpha^\prime t)}
{_4F_3}\left({
{\frac{3}{2}-\alpha ,1-\beta ,-\alpha -2 \nu ,1+\alpha^\prime t} \atop {1-\alpha ,\frac{1}{2}-\alpha -2 \nu ,2+\alpha^\prime t }}\bigg| y\right)
\\
&& +
\frac{\Gamma\!\left(\frac{1}{2}+2\nu \right)\Gamma(-\alpha)  \Gamma(\beta)\, y^{1+\alpha(t)}
}{
 \Gamma\!\left(\frac{1}{2}\right)\Gamma(1+2\nu)\Gamma(\beta -\alpha)\left(1+\alpha(t)\right)}
{_4F_3}\left({
{-2\nu ,\frac{3}{2},1+\alpha -\beta ,1+\alpha(t)} \atop {\frac{1}{2}-2\nu ,1+\alpha ,2+\alpha(t)}}\bigg| y\right)
\Bigg].
\nonumber
\end{eqnarray}
The forward-like functions here look a bit more intricate
compared to the generic dual parametrization Ansatz
(\ref{ansatz-generic}).
However, for given $\nu$ values
these expressions reduce to sums of the
$\Gamma$-function ratios, which are further simplified at
$t=0$.
The three first forward-like functions
(\ref{KMreexpressed})
and the corresponding GPD quintessence function
$N(y,t)=\sum_{\nu=0}^2 y^{2\nu} Q_{2\nu}(y,t)$
are presented  in the lower right panel of
Fig.~\ref{fig-gpd3D}
for
$t=0$
as functions of
$y$.
As one realizes the
$\nu=0$
(dash-dotted) and
$\nu=1$
(dashed)
forward-like functions
turn to be positive and are roughly of the same size while the
$\nu=2$
one (dotted) is negative. Therefore, the GPD quintessence
$N(y,t)$
is smaller than the
$\nu=0$ contribution. This ensures that the normalization of the DVCS cross section
is correctly described and that the skewness ratio $r^q$ remains almost unchanged under QCD evolution.

The parton density
(\ref{Q_02H(x,0,t)})
can be recovered from the convolution of
$Q_0$
with the  kernel
(\ref{Forward_limit_kernel}).
Alternatively, the inverse Mellin transform of the
$\nu=0$
GPD moments
(\ref{KM10-ansatz})
yields the equivalent
result, which turns out to be rather simple
\begin{eqnarray}
H(x,\eta=0,t)&\!\!\!=\!\!\! & M_2 \frac{\Gamma(3-\alpha +\beta)\,x^{-\alpha(t)}}{\Gamma(2-\alpha)\Gamma(\beta)}\, \mbox{\boldmath $ \beta$}(t)  \Bigg[
\frac{ \Gamma(1+\alpha^\prime t) \Gamma(\beta)}{\Gamma(1+\alpha^\prime t+\beta)}
-\frac{x^{1+ \alpha^\prime t} }{1+\alpha^\prime t}\, {_2F_1}\!\!\left(\!{1-\beta ,1+\alpha^\prime t \atop 2+\alpha^\prime t}\bigg|x\!\right)
\Bigg].
\nonumber\\
\end{eqnarray}
For
$t=0$
it reduces to the well known building block for PDFs
$$
H(x,\eta=0,t=0)= M_2 \frac{\Gamma(3-\alpha +\beta)}{\Gamma(2-\alpha) \Gamma(1+\beta)} x^{-\alpha }  (1-x)^{\beta }\,,
$$
{\it cf.}
(\ref{H^{(+)}-nfPDF})
of the toy Ansatz
(\ref{ansatz-generic}).
Although the analytic formulae for the forward-like functions and the corresponding
$t$-dependent PDFs look a bit different for the toy model
(\ref{ansatz-generic})
and the Ansatz used for {\em KM10}
fit, we might state that the differences of the both Ans\"atze, caused by the different normalization in
(\ref{HjJ2B-toy}),
are marginal.

\section{Partial wave amplitudes from double distributions}
\label{sec:dPWAsfromDDs}
\setcounter{equation}{0}

A reparametrization procedure, allowing to map any particular GPD to the forward-like function as it appears in the dual parametrization,
was proposed in Ref.~\cite{SemenovTianShansky:2008mp}. This procedure is based on the Taylor expansion of GPDs in the vicinity of $\eta=0$ and
an example was given in Ref.~\cite{Polyakov:2008aa}, where
several first forward-like functions reexpressing the RDDA within the dual parametrization approach were computed.
In this section we propose a complementary method that is based on the evaluation of dPWAs (\ref{B_{nl}}), (\ref{HjJ2B})
from a DD. We provide the desired map  in terms of a convolution integral, where the integral kernel is given in a closed form.
The inverse Mellin transform (\ref{HjJ2Q}) also allows the reconstruction of the set of the forward-like function for
recasting the DD representation in the dual parametrization framework. This provides
a useful and independent cross check of our method.

To be more general, we adopt the following DD representation of the charge even GPD
\begin{eqnarray}
&&
H^{(+)}(x,\eta) =\frac{1+a(1-2x)}{1+a^2}\! \int_0^1\!\!dy\!\int_{-1+y}^{1+y}\!dz\,  \delta(x-y-z\eta)  f(y,z|a) + \frac{D(x/|\eta|;a) }{2} - \{x\to -x\}
\nonumber \\ &&
\label{DD_with_parameter_a}
\end{eqnarray}
in terms of the DD $f(y,z,t|a)$  and the  $D$-term
$D(x/|\eta|;a)=\theta(|x|\le |\eta|) d(x/|\eta|;a)$.
Specifying the parameter
$a$
allows us to switch between several popular choices of the DD-representation:
\bi
\item $a=0$ corresponds to the `standard' DD  parametrization that requires a $D$-term to complete polynomiality \cite{Polyakov:1999gs}.
This DD-representation is commonly used by model builders.

\item $a=+1$  corresponds to the so-called one-component DD
representation for a pion GPD \cite{Belitsky:2000vk}. It has been also adopted to the nucleon GPD \cite{Radyushkin:2013hca} and was also numerically studied in \cite{Mezrag:2013mya}.
\item  Finally, the $a=-1$  produces a $x$ factor as it appears for instance for a nucleon GPD $E$ in a diquark model \cite{Hwang:2007tb},
which is on more general ground suggested by positivity constraints.
\ei
We emphasize that in the $a\neq 0$ cases polynomiality is completed without a $D$-term and that all these representation are equivalent.
They can be mapped to each other by an integral transformation of the DD
(see Refs.~\cite{Belitsky:2000vk,Gabdrakhmanov:2012aa,Mueller:2014hsa} for examples),
which can be understood on a more general ground as a `gauge' transformation \cite{Teryaev:2001qm}.
Note that the $D$-term can be also presented as an  addenda to
$f(y,z,t)$
that is proportional to
$\delta(y)$,
\be
\frac{1+a(1-2x)}{1+a^2} f(y,z |a)\quad \Rightarrow \quad \frac{1+a(1-2x)}{1+a^2} f(y,z |a) + \frac{x\, D(z|a)}{2z}  \delta(y).
\label{Addenda_DD}
\ee

By employing the definition of the PW-amplitudes
(\ref{B_{nl}}),
(or alternatively the
${\rm SO}(3)$
partial wave decomposition
(\ref{H_n(eta,t)-SO(3)})),
we work out the following expressions for the GPD moments for the integer values
of the conformal spin and cross channel angular momentum:
\begin{eqnarray}
&&
H^{(+)}_{n+2\nu,n+1}(a) =
\frac{\Gamma(3+n+2\nu)\Gamma\!\left(\frac{3}{2}+n\right)}{2^{2\nu} \Gamma\!\left(\frac{5}{2}+n+2\nu\right)\Gamma(2+n)}
\int_{-1}^1\!d\omega\, \frac{3+2n}{2} P_{n+1}(\omega)
\\
&&\times\int_0^1\!dy\int_{-1+y}^{1+y}\!dz\,
\frac{(1+a)\omega-2a(\omega y+z)}{1+a^2}f(y,z |a)
\frac{(3+2n+4\nu)C_{n+2\nu}^{3/2}(\omega y+z)}{2(1+n+2\nu)(2+n+2\nu)} + \delta_{n+1,0}\, D_{2\nu,0}(t) .
\nonumber
\label{Moments_DD}
\end{eqnarray}

Now we need to specify the analytic continuation of the GPD moments
(\ref{Moments_DD})
to the complex values of $n=j$.
Using the Rodrigues formula for the Legendre polynomials
$$
 P_l(\omega) = \frac{(-1)^l}{2^l l!}\, \frac{d^l}{d\omega^l}\, \left(1-\omega^2\right)^l
$$
and the definition of the Gegenbauer polynomials in terms of hypergeometric functions,
$$
 C_{n}^{3/2}(x)= \frac{(n+1)(n+2)}{2}\,
{_2F_1}\!\!\left({-n,3+n+2\atop 2}\bigg|\frac{1-x}{2}\!\right),
$$
allows us to reshuffle the derivatives
w.r.t.~$\omega=1/\eta$
by means of partial integration.
Consequently, the GPD moments can be written as
\begin{eqnarray}
\label{DD2H^{(+)}_{j+2nu,j+1}}
H^{(+)}_{j+2\nu,j+1}(a) \!\!\!&\!\!\!=\!\!\!&\!\!\! \int_0^1\!dy\!\!\int_{-1+y}^{1-y}\!dz\,
\frac{y^{j}(1+a-2ay)}{1+a^2}  f(y,z|a )\!\left[
 1+j+  \frac{ y\vec{\partial}}{\partial y}-
\frac{2 a y}{1+a-2ay}\frac{z\vec{\partial}}{\partial z}
\right]\!
 {\cal K}_{j+2\nu,j+1}^{-1}(y,z)
\nonumber\\
&&\!\!\! + \delta_{j+1,0}\, D_{2\nu-1,0}(t)\,,
\end{eqnarray}
where the integral kernel
turns to be a polynomial in
$y$
and
$z$.
This kernel is given by acting with the differential operator on
the Gegenbauer polynomials
\begin{eqnarray}
\label{ K_{j+2nu,j}^{-1}(y,z)-1}
{\cal K}_{j+2\nu,j+1}^{-1}(y,z) &\!\!\!=\!\!\!&
\frac{(1+2 \nu)_j (2+j)_\nu}{\Gamma(2+j)\left(\frac{3}{2}+j+\nu\right)_\nu}
\nonumber\\
&&\times\int_{-1}^{1}\!d\omega\,
\frac{\left(2+j\right)_{2+j} \left(1-\omega^2\right)^{1+j}}{2^{3+2 j}\,\Gamma(2+j)}\,
{_2F_1}\!\!\left({-2\nu,3+2j+2\nu\atop 2+j }\bigg|\frac{1-\omega y-z}{2}\!\right).
\nonumber\\
\end{eqnarray}
The transformation
(\ref{DD2H^{(+)}_{j+2nu,j+1}}), (\ref{ K_{j+2nu,j}^{-1}(y,z)-1})
provides the analytic continuation of the GPD moments in
$j$ for the regular part of the GPD.
For the
$j=-1$
Kronecker delta contribution induced by the $D$-term part
it can be calculated from a
$\delta(y)$ addenda
(\ref{Addenda_DD})
to DD
by convolution with the kernel $z \frac{\partial}{\partial z} K^{-1}_{2\nu-1,0}(y=0,z) $,
which  explicitly  reads as
\begin{eqnarray}
D_{2\nu-1,0}(t) &\!\!\! =\!\!\!&
\frac{\Gamma\!\left(\frac{1}{2}\right)\Gamma(2+2\nu)}{2^{2\nu} \Gamma\!\left(\frac{1}{2}+2\nu\right)}
\int_{-1}^1\! dz\,  D(z,t)\, {_2F_1}\!\!\left({1-2\nu,2+2\nu\atop 2 }\bigg|\frac{1-z}{2}\!\right).
\end{eqnarray}

A convenient double integral representation for the kernel
(\ref{ K_{j+2nu,j}^{-1}(y,z)-1})
follows
from the definition of the
${_2 F_1}$
hypergeometric function in terms of the analytical regularized integral,
\begin{eqnarray}
\label{K_{j+2nu,j}^{-1}(y,z)-integral}
 {\cal K}_{j+2\nu,j+1}^{-1}(y,z) &\!\!\!=\!\!\!& \frac{(1+2 \nu)_j (2+j)_\nu}{\Gamma(2+j)\left(\frac{3}{2}+j+\nu\right)_\nu}\,
  \frac{\sin(\pi  j)\,(2+j)_{2+j}}{\pi\,\Gamma(2+j) (2+j+2 \nu)_{1+j} }\, \frac{2^{1+2 b}\Gamma(1+2 \nu)}{\Gamma(2+2 b+2 \nu)}
\\
 &&\qquad\qquad
\times \frac{\partial^{2b+1}}{\partial z^{2b+1}}\int_0^1\!du\!\!\int^{(1)}_0\!dv
\left(u\overline{u}\right)^{1+j} \frac{v^{2(1+\nu +j)-2b-1} }{\overline{v}^{2(1+\nu) +j} }
 \left[1-v\,\frac{1-(\overline{u}-u )y-z }{2} \right]^{2\nu}.
 \nonumber
\end{eqnarray}
Note that we have introduced an arbitrary parameter
$b\ge -1/2$.
In what follows it will turn out convenient to equate it to the $b$ parameter of the profile function
$h^{(b)}(z,y)$
of the factorized RDDA.


To specify the polynomials, we are dealing with,
we expand the integrand in the vicinity of
$z=0$
and employ a quadratic variable transformation.
In this way we work out the closed form of the kernel
${\cal K}_{j+2\nu,j+1}^{-1}(y,z)$
in terms of the rather cumbersome Appell hypergeometric function ${F_4}$,
which finally provides us the finite double sum
\be
{\cal K}_{j+2\nu,j+1}^{-1}(y,z) &\!\!\!=\!\!\!&
\frac{(-1)^\nu 2^{-2 \nu} (2+j)_{2 \nu -1}}{\Gamma(\nu+1) \left(\frac{3}{2}+j+\nu\right)_\nu}
{F_4}(-\nu,\frac{3}{2}+j+\nu,\frac{5}{2}+j,\frac{1}{2},y^2,z^2)
\nonumber \\
&\!\!\!=\!\!\!&
\frac{ (2+j)_{2 \nu -1}}{\left(\frac{3}{2}+j+\nu\right)_\nu}
\sum_{m=0}^\nu \sum_{n=0}^{\nu-m} \frac{(-1)^{\nu-m-n}\,  y^{2m} z^{2n}}{m! n! (\nu -m-n)!\,  2^{2\nu}}\,
\frac{ \left(\frac{3}{2} + j+\nu\right)_{m+n}
}{
\left(\frac{5}{2} + j\right)_{m}\left(\frac{1}{2}\right)_{n}},
\label{K_{j+2nu,j}^{-1}(y,z)-sum}
\ee
where
$(k)_l \equiv \frac{\Gamma(k+l)}{\Gamma(l)}$ is the Pochhammer symbol.

In order to illustrate the application of the above reparametrization procedure
let us map the RDDA into the space of GPD moments.
Taking care on the forward limit
the factorizable RDDA  reads in an arbitrary `gauge' as
$$
f(y,z|a) = h^{(b)}(y,z|a) q(b)= \frac{\left(1+a^2\right)}{[1+a(1-2y)](1-y)}\, \frac{\Gamma\!\left(\frac{3}{2}+b\right) }{\Gamma\!\left(\frac{1}{2}\right) \Gamma(1+b)}
\left(1-\frac{z^2}{(1-y)^2}\right)^b q(y).
$$
We adopt the PDF  parametrization
$$
q(y,t)=\frac{M_2 \Gamma(3-\alpha +\beta)}{\Gamma(2-\alpha)\Gamma(1+\beta)}  y^{-\alpha} (1-y)^\beta\,
$$
with the first Mellin moment normalized to the momentum fraction
$M_2$
carried by quarks.

Applying the differential operator on the double sum
(\ref{K_{j+2nu,j}^{-1}(y,z)-sum})
and changing to the integration variable
$w= z/(1-y)$,
the transformation
(\ref{DD2H^{(+)}_{j+2nu,j+1}})
can be written as
\begin{eqnarray}
\label{DD2H^{(+)}_{j+2nu,j+1}-RDDA}
H^{(+)}_{j+2\nu,j+1} &\!\!\!=\!\!\!& \frac{M_2 \Gamma(3-\alpha +\beta)}{\Gamma(2-\alpha)\Gamma(1+\beta)}\,
\frac{\Gamma\!\left(\frac{3}{2}+b\right) }{\Gamma\!\left(\frac{1}{2}\right) \Gamma(1+b)}
\int_0^1\!dy\!\!\int_{-1}^{1}\!dw\,   y^{j-\alpha }(1-y)^{\beta} \left(1-w^2\right)^b
\nonumber\\
&&\times
\sum_{m=0}^\nu \sum_{n=0}^{\nu-m} \frac{(-1)^{\nu-m-n}\,  y^{2m} (1-y)^{2n} w^{2n}}{m! n! (\nu -m-n)!\,  2^{2\nu}}\,
\frac{(1+j)_{2 \nu} \left(\frac{3}{2} + j+\nu\right)_{m+n}
}{
\left(\frac{5}{2} + j\right)_{m}\left(\frac{3}{2}+j+\nu\right)_\nu\left(\frac{1}{2}\right)_{n}}
\nonumber\\
&&\times
\frac{1}{1+j}\left[1+\frac{2 m}{1+j}- \frac{2 n}{1+j}\,\frac{2 a y}{1+a-2 a y}
\right]\,.
\end{eqnarray}
For the three popular choices  $a\in\{-1,0,1\}$ the result can be straightforwardly evaluated in
terms of a double sum,
\begin{eqnarray}
H^{(+)}_{j+2\nu,j+1} &\!\!\!=\!\!\!&  \frac{M_2 (2-\alpha)_{j-1} }{(3-\alpha +\beta)_{j-1}}
\sum_{m=0}^\nu \sum_{n=0}^{\nu-m} \frac{(-1)^{\nu-m-n}\, 2^{-2\nu}}{m! n! (\nu -m-n)!}\,
\frac{(1+j)_{2 \nu}\left(\frac{3}{2} + j+\nu\right)_{m+n}
}{
\left(\frac{5}{2} + j\right)_{m}\left(\frac{3}{2}+j+\nu\right)_\nu}\,
\frac{(1+\beta)_{2n} }{\left(\frac{3}{2}+b\right)_n}
\nonumber\\
&&\times \frac{(1-\alpha+j)_{2m}}{(2-\alpha +\beta+j)_{2 m+2 n}}
\left[1+2\frac{m -a\, n}{1+j} - a(a+1) \frac{n(1+j-\alpha +\beta +2 m-2 n)}{(1+j)(\beta +2 n)}  \right].
\nonumber\\
\label{RDDA_moments_reparametrized}
\end{eqnarray}
For the lowest two
$\nu$
values
$\nu=0,\,1$
the explicit results read, {\it e.g.}, for the $a=1$ `gauge' as follows
\begin{eqnarray}
H^{(+)}_{j,j+1}&\!\!\!=\!\!\!&\frac{M_2 (2-\alpha)_{j-1}}{(3-\alpha +\beta)_{j-1}}\\
H^{(+)}_{j+2,j+1}&\!\!\!=\!\!\!&H^{(+)}_{j,j+1}\,\frac{4+2j}{5+2 j}
\left[
\frac{1+j}{2}-\frac{(3+j) (1+j-\alpha )_2}{2 (2+j-\alpha +\beta )_2}
-\frac{(5+2 j) (1+\beta ) (2 \alpha +\beta +j \beta )}{2(3+2 b) (2+j-\alpha +\beta )_2}\right]\,.
\nonumber
\end{eqnarray}
To make link to the dual parametrization framework we note that, by the use of the inverse Mellin transform
(\ref{HjJ2Q}),
we obtain  the usual expression for the forward-like function
$Q_0$
and recover the result of
\cite{Polyakov:2008aa}
for the forward-like function
$Q_2$.

Finally, let us comment on the functional form of the moments
(\ref{RDDA_moments_reparametrized})
as they arise from the RDDA. Generally, we expect that for arbitrary
$\alpha$, $\beta$,
and
$b$
parameters  the GPD moments can not be expressed by a single higher order
hypergeometric function, however, for  specific parameter values a simplification occurs. \\

\noindent
\textbullet\quad {$b\to \infty$ limit}

\noindent
The limiting case $b\to \infty$, corresponding to a $\delta(z)$  (or $\delta(w)$) function in the DD-representation, can be trivially treated
in the representation (\ref{DD2H^{(+)}_{j+2nu,j+1}-RDDA}). Since only the $n=0$ term in the double sum contributes, the summation and $y$-integration
can be straightforwardly performed and the `gauge' independent result reads in terms of $_4F_3$ functions as follows
\begin{eqnarray}
H^{(+)}_{j,j+1}&\!\!\!\stackrel{b\to\infty}{=}\!\!\!&
\frac{ (2-\alpha)_{j-1}}{(3-\alpha +\beta)_{j-1} }
\frac{(-1)^{\nu } 2^{-2 \nu }  (1+j)_{2 \nu}}{ \Gamma(1+\nu) \left(\frac{3}{2}+j+\nu\right)_\nu}
\\
&&\phantom{\frac{ (2-\alpha)_{j-1}}{(3-\alpha +\beta)_{j-1} }}\times\left[1+ \frac{2}{1+j}\frac{\vec{d}}{dy} \right]
{_4F_3}\left({
-\nu,\frac{3}{2}+j+\nu  ,\frac{1}{2}+\frac{j}{2}-\frac{\alpha }{2},1+\frac{j}{2}-\frac{\alpha }{2}
\atop
\frac{5}{2}+j,1+\frac{j}{2}-\frac{\alpha }{2}+\frac{\beta }{2},\frac{3}{2}+\frac{j}{2}-\frac{\alpha }{2}+\frac{\beta }{2}
}\bigg|y\right)\Bigg|_{y=1}\,.
\nonumber
\end{eqnarray}
Here the derivative w.r.t.~$y$ provides, apart from a prefactor, the shift by one unit of all parameters in the hypergeometric function,
$$
\frac{d^n}{dx^n}\, {_pF_q}\left( {a_1,\cdots, a_p \atop b_1,\cdots, b_q }\big| x\right) =  \frac{(a_1)_n\cdots (a_p)_n}{(b_1)_n\cdots (b_q)_n}\,
{_pF_q}\left( {a_1+n,\cdots, a_p+n \atop b_1+n,\cdots,b_q+n}\big| x\right).
$$

\noindent
\textbullet\quad{integer $\beta$ or $b$ values}

\noindent
For integer $\beta$ values
the $_4F_3$ can be reduced to a sum of $_2F_1$ functions, {\it i.e.}, finally one gets a finite sum of $\Gamma$-function ratios.
Also for integer  $b$ values the GPD moments of the RDDA can be reduced for arbitrary $\nu$ values to finite sums. This follows most easily from
the integral representation of the kernel
(\ref{K_{j+2nu,j}^{-1}(y,z)-integral}),
where the arbitrary $b$-parameter is identified with the profile parameter of RDDA. Partial integration
yields only boundary terms at
$z=\pm (1-y)$,
at which the auxiliary double integral can be expressed in terms of $_2F_1$ functions,
whose Mellin moments w.r.t.~$y$ yield $_3F_2$ functions.

\noindent
\textbullet\quad integer $b-\alpha$ value

\noindent
Also it is worth mentioning that from
(\ref{RDDA_moments_reparametrized})
we notice that when choosing
$b=\alpha+M$, $M$ - integer the leading
$j=\alpha-1$ singularity is present only in the dPWAs $H^{(+)}_{j+2 \nu,j+1}$
with
$2\nu \le 2M$. In particular, this means that it suffices to account just a
finite number of conformal PWs to reproduce the
small-$\xi$  asymptotic behavior of the imaginary part of elementary amplitude
\be
\left. {\rm Im} {\cal H}^{\rm S}(\eta=\xi) \right|_{{\rm RDDA} \, a=0} \sim
\frac{2^{2b+1-\alpha}}{\xi^\alpha}
\frac{\Gamma(\frac{1}{2}) \Gamma(b+ \frac{3}{2}) \Gamma(1+b-\alpha)}{\Gamma(2+2b-\alpha)}
\label{ImA_DD_small_xi_asymp}
\ee
within RDDA. Instead, for arbitrary real
$b \ge - \frac{1}{2} $ one has to take into account an infinite number of conformal PWs
to reproduce the RDDA
small-$\xi$  asymptotic behavior of
${\rm Im} {\cal H}^{\rm S}(\xi)$.
However, the contribution of the conformal PWs with $2 \nu > 2M$, where $M$ is the least integer
$b < \alpha+  M$ turn to be suppressed by $(\alpha+M-b)$ factor and by the arithmetic combinatorial factors.

\section{Conclusions}
\label{sec:summary}

The problem of building up a phenomenological GPD parametrization capable of describing the whole set of the
present day experimental data can be in principle addressed within any GPD representation:
double distribution representation, dual parametrization, or Mellin-Barnes integral representation.
The main ingredients for this issue are flexible phenomenological GPD models and the development
of the state-of-the-art-level formalism allowing to systematically account for the next-to-leading order
and  kinematical twist-four corrections \cite{Braun:2011zr,Braun:2011dg,Braun:2012bg,Braun:2012hq,Braun:2014sta}.
The global fitting procedure \cite{Kumericki:2007sa,Kumericki:2009uq,Mueller:2013caa}
developed within the Mellin-Barnes integral representation framework
represents the up-to-present-date most consecutive approach to challenge the GPD data description.

The equivalence of the cross channel ${\rm SO}(3)$
partial wave expansion implementation within the Mellin-Barnes integral GPD representation
and the dual parametrization of GPDs was taken for granted by us already for some time,
however, it was certainly not widely realized. In this study we explicitly demonstrated the full mathematical
equivalence of these two GPD representations.
\bi
\item We established the explicit mapping
(\ref{HjJ2Q})
between the forward-like functions
of the dual parametrization and the set of double partial wave amplitudes employed within the Mellin-Barnes integral approach.
\item The dual parametrization convolution kernels were
 reexpressed within the Mellin-Barnes integral representation,
 {\it cf.} Eq.~(\ref{K_{2nu}(x,eta|y)-MB}).
\item To illustrate the equivalence of the two representations  we considered in details several special  limiting cases:
$\eta \to 0$
limit
(Sec.~\ref{ssec:fwd}),
GPD on the cross-over line
$x=\eta$
(Sec.~\ref{sec-cases_cross-over}),
$\eta \to 1$ limit
(Sec.~\ref{ssec:eta_1})
and the `low energy' limit $\eta \to \infty$
(Sec.~\ref{ssec:D_term}).

\item We reexpress the successful {\em KM10} model in the dual parametrization framework
and present explicit analytic expressions
(\ref{KMreexpressed})
for the set of the corresponding forward-like functions
$Q_{2\nu}(y,t)$.

\item We also address the reparametrization procedure allowing to
recast the double distribution  representation of GPDs in the Mellin-Barnes
integral framework. We presented for the first time a closed formula
(\ref{DD2H^{(+)}_{j+2nu,j+1}})
in terms of Appell`s
$F_4$
function
(\ref{K_{j+2nu,j}^{-1}(y,z)-sum})
that allows to map
the double distribution into the space of double partial wave
amplitudes with complex conformal spin%
\footnote{It is interesting to remark that double distributions
can be obtained from GPDs by the inverse Radon transform where, however, the GPD support has to be extended into the unphysical region.
Therefore, the  mathematical aspects of invertibility of the transformation (\ref{DD2H^{(+)}_{j+2nu,j+1}}) represent an interesting problem.
}. Consequently, with help of
(\ref{HjJ2Q})
one can also map a double distribution numerically into the
space of forward-like functions as they appear in the dual parametrization.
\ei


The equivalent Mellin-Barnes integral representation and the dual parametrization of GPDs
can be also seen as somewhat complementary. What looks most simple in one representation
turns out to be a bit more complicated in the other one and vice versa. This allows to enlighten
the physical content of GPDs form different perspectives.


In particular, we would like to stress several points with respect to the discussion of
the physical meaning of the GPD quintessence function.
To the leading order accuracy, the GPD quintessence function and the
GPD on the cross-over line are the two pseudonyms for
the imaginary part of the DVCS amplitude. They are related by the Abel integral transformation.
The Mellin
moments of the GPD quintessence function
(\ref{a_J(t)_Dual}),
(\ref{a_J=0(t)_Dual})
express the partial wave amplitudes corresponding to the definite value of the
$t$-channel angular momentum. This property is equivalent to the so-called Froissart-–Gribov projection
(\ref{Froissart-–Gribov}),
(\ref{Froissart-–Gribov_J=0}
that allows to compute the corresponding partial wave amplitudes by convolutions of the imaginary part of the DVCS amplitude with
the  Legendre functions of the second kind.

Another important quantity is the
$D$-form factor,
the subtraction
constant in the fixed-$t$ dispersion relation for the elementary DVCS amplitude.
The freedom in fixing this subtraction constant is related to the
$J=0$ fixed pole contribution  to the charge even GPD
$H^{(+)}$. As in our previous studies, we conclude that the $D$-form factor may
receive a $J=0$ fixed pole contribution
(\ref{D^{f.p.}(t)})
from the forward-like functions
$Q_{2 \nu}(y,t)$
with
$\nu \ge 1$.
Therefore, in principle, the $J=0$ fixed pole contribution into the
$J=0$ partial wave (\ref{Analytic_J=0_PW})
may depend on the external kinematics.
This is at variance with the claim of
Refs.~\cite{Brodsky:2008qu,Szczepaniak:2007af}
that
the
$J=0$
fixed pole contribution into the DVCS amplitude represents the universal quantity
independent on the external kinematics and arises solely from the local two-photon interaction contribution
proportional to the analytically regularized inverse Mellin momentum of the corresponding
$t$-dependent PDF.
This statement is in fact only valid once the external ``analyticity principle'' requiring
$D^{\rm f.p.}(t)=0$
is attained and the $D$-form factor is unambiguously fixed from the known
GPD quintessence function
$N(y,t)$
and forward-like function
$Q_0(y,t)$
assuming the analytic regularization for the potentially divergent integrals
(see Eq.~(\ref{D_ff_dual_parametrization_regularized}))%
\footnote{Or, equivalently, the $D$-form factor is computed with the help of the GPD sum rule
(\ref{Inverse_momentum_SR})
from the knowledge of the GPD on the cross-over line and the corresponding $t$-dependent PDF.}.
Unfortunately, the required analyticity assumption lacks solid theoretical ground and can be
verified  from experimental measurements only in a model dependent manner.
Therefore, it remains unclear to us how much
biased such studies will be at the very end. The present world data set of DVCS measurements allows only to
determine the $D$-form factor on a qualitative level and in present global DVCS fits its order of magnitude and sign coincide
with theoretical expectations \cite{Goeke:2001tz,Goeke:2007fp,Pasquini:2014vua}.

\subsection*{Acknowledgements}
D.M. and M.V.P. are indebted to the members of the IFPA group for the warm hospitality during their stays at the University of Li{\`e}ge,
where this project has been staged and finalized. M.V.P. acknowledges support of RSF grant 14-22-0281.

\end{document}